\documentclass[12pt, preprint,twoside]{aastex}
\usepackage{geometry}               
\usepackage{graphicx}
\usepackage{amssymb}
\usepackage{epstopdf}
\usepackage{fullpage}
\usepackage{natbib}
\usepackage{subfigure}
\usepackage{epsfig}
\usepackage{multirow}
\usepackage[figuresright]{rotating}
\usepackage{amsmath}

\renewcommand{\&}{and}

\setcitestyle{round,aysep={,},yysep={;}}
\title{The Effect of Orbital Configuration on the Possible Climates and Habitability of Kepler-62f}
\author{Aomawa L. Shields\altaffilmark{1,2,5},
Rory Barnes\altaffilmark{3,5},
Eric Agol\altaffilmark{3,5},\\
Benjamin Charnay\altaffilmark{3,5},
Cecilia Bitz\altaffilmark{4,5}
Victoria S. Meadows\altaffilmark{3,5},\\
}

\altaffiltext{1}{\bf{Corresponding author: Aomawa Shields, Department of Physics and Astronomy, University of California, Los Angeles, Box 951547, Los Angeles, CA 90095-1547. ashields@astro.ucla.edu.}}
\altaffiltext{2}{NSF Astronomy and Astrophysics Postdoctoral Fellow, UC President's Postdoctoral Program Fellow, Department of Physics and Astronomy, University of California, Los Angeles and Harvard-Smithsonian Center for Astrophysics}
\altaffiltext{3}{Department of Astronomy and Astrobiology Program, University of Washington}
\altaffiltext{4}{Department of Atmospheric Sciences, University of Washington}
\altaffiltext{5}{NASA Astrobiology Institute's Virtual Planetary Laboratory}

\begin{document}
\begin{abstract}
As lower-mass stars often host multiple rocky planets, gravitational interactions among planets can have significant effects on climate and habitability over long timescales. Here we explore a specific case, Kepler-62f \citep{Borucki2013}, a potentially habitable planet in a five-planet system with a K2V host star. \emph{N}-body integrations reveal the stable range of initial eccentricities for Kepler-62f is $0.00 \leqslant e \leqslant 0.32$, absent the effect of additional, undetected planets. We simulate the tidal evolution of Kepler-62f in this range and find that, for certain assumptions, the planet can be locked in a synchronous rotation state. Simulations using the three-dimensional (3-D) ``Laboratoire de M\'et\'eorologie Dynamique" (LMD) Generic global climate model (GCM) indicate that the surface habitability of this planet is sensitive to orbital configuration. With 3 bars of CO$_2$ in its atmosphere, we find that Kepler-62f would only be warm enough for surface liquid water at the upper limit of this eccentricity range, providing it has a high planetary obliquity (between 60$^\circ$ and 90$^\circ$). A climate similar to modern-day Earth is possible for the entire range of stable eccentricities if atmospheric CO$_2$ is increased to 5-bar levels. In a low-CO$_2$ case (Earth-like levels), simulations with version 4 of the Community Climate System Model (CCSM4) GCM and LMD Generic GCM indicate that increases in planetary obliquity and orbital eccentricity coupled with an orbital configuration that places the summer solstice at or near pericenter permit regions of the planet with above-freezing surface temperatures. This may melt ice sheets formed during colder seasons. If Kepler-62f is synchronously rotating and has an ocean, CO$_2$ levels above 3 bars would be required to distribute enough heat to the night side of the planet to avoid atmospheric freeze-out and permit a large enough region of open water at the planet's substellar point to remain stable. Overall, we find multiple plausible combinations of orbital and atmospheric properties that permit surface liquid water on Kepler-62f. \end{abstract}
\maketitle
\pagebreak

%\linenumbers
\section{Introduction}
	NASA's \emph{Kepler} mission \citep{Borucki2006}, launched in 2009, identified more than 4700 transiting planet candidates\textemdash over 1000 of which have been confirmed as planets\textemdash in its first five years of observations\protect\footnotemark{}. Recent statistical surveys estimate $\sim$40\% of planetary candidates to be members of multiple-planet systems \citep{Rowe2014}. Analyses also suggest a low false-positive probability of discovery, indicating that the clear majority of these multiple-planet candidates are indeed real, physically associated planets \citep{Lissauer2012b, Lissauer2014}. Kepler data also indicate that smaller stars are more likely to host a larger number of planets per star \citep{Swift2013}, and smaller planets are more abundant around smaller stars \citep{Howard2012, Mulders2015}. \footnotetext{http://kepler.nasa.gov/ as of February 12, 2016}These statistical data suggest that systems of multiple small planets orbiting low-mass stars are a major new planetary population, and one that may be teeming with habitable worlds \citep{Anglada2013}.

	One of the systems of particular interest for habitability orbits Kepler-62 [Kepler Input Catalog (KIC) 9002278, Kepler Object of Interest (KOI) 701], a K2V star. With a mass of 0.69M$_\odot$ and a radius of 0.63R$_\odot$, Kepler-62 hosts five planets\textemdash Kepler-62b-f\textemdash with orbital periods ranging from $\sim$5 to $\sim$267 days, and radii from 0.54 to 1.95R$_\oplus$ \citep{Borucki2013}. The inclinations of all five planets are edge-on ($i$$\sim$89-90$^\circ$). However, neither their orbital eccentricities nor their planetary obliquities are constrained. The two outermost planets, Kepler-62e (1.61R$_\oplus$) and Kepler-62f (1.41R$_\oplus$) receive 120\% and 41\% of the amount of flux that Earth receives from the Sun, respectively. Kepler-62e, with an equilibrium temperature of $\sim$270 K without an atmosphere, is likely to be too hot (with an atmosphere) to have liquid water on its surface, although a significant amount of cloud cover could reflect incident radiation and cool a planet, if it is synchronously rotating \citep{Yang2013, Yang2014}. In contrast, Kepler-62f sits near the outer edge of the habitable zone, with a relatively low incoming stellar insolation (hereafter ``instellation"). However, the planet could avoid freezing with a sufficient greenhouse effect \citep{Kaltenegger2013}, although there are few measured planetary characteristics that could help constrain whether habitability is likely. 

	Based on planet size and instellation, Kepler-62f is the most likely candidate for a habitable world in this system. While Kepler-62f has a radius of 1.41 R$_\oplus$ \citep{Borucki2013}, its mass is unknown, leaving its density also unconstrained. Recent statistical surveys of a population of \emph{Kepler} planets with known masses (from companion RV surveys) found that volatile inventory increases with planet radius \citep{Marcy2014a, Weiss2014}. However, the most recent bayesian analysis of \emph{Kepler} planets with RV follow-up data found that planets with radii $\leqslant$ 1.6R$_\oplus$ are likely sufficiently dense to be rocky (composed of iron and silicates), though this study was limited to close-in planets with orbital periods less than 50 days \citep{Rogers2015}.  
		
	There are many planetary properties that affect the presence of surface liquid water and perhaps life, and in the absence of sufficient observational data, many of the effects of these properties can only be explored using a global climate model (GCM). Previous work on the climate modeling of Kepler-62f found that 1.6-5 bars of CO$_2$ would yield surface temperatures above the freezing point of liquid water, depending on the mass and surface albedo of the planet \citep{Kaltenegger2013}. However, this work was done using a 1-D radiative-convective atmospheric code, and did not include a treatment of clouds beyond a scaling of surface albedo. Additionally, the effect of the orbital architecture and evolution of this multiple-planet system on the climate of Kepler-62f was not explored. The eccentricity of a planet could be pumped to high values in the presence of additional companions in the system \citep{Mardling2007, Correia2012}, and the eccentricity of Kepler-62f is poorly constrained. The effect of different rotation rates on atmospheric circulation was also not examined. Changes in planetary obliquity \citep{Ward1974, Williams1975, Williams1997, Dobrovolskis2013} and eccentricity \citep{Berger1993, Berger2006} will affect the seasonality and average instellation received by a planet throughout its orbit, respectively. Simulations have found that a large moon around Kepler-62f could be long lived over timescales necessary for biological evolution, and could help stabilize the planet's obliquity and climate  \citep{Sasaki2014}, as the Moon has for the Earth \citep{Laskar1993}. However, large moons may not be required to stabilize a planet's obliquity over long timescales, given that variations in a planet's obliquity without a moon would be constrained \citep{Lissauer2012a} and slowly evolving \citep{Li2014}. Large moons could even be detrimental at the outer edge of the habitable zone \citep{Armstrong2014}. Additionally, planetary rotation rate has been shown to affect atmospheric circulation \citep{Joshi1997, Merlis2010, Showman2011a, Showman2011b, Showman2013}. Quantifying the effect of orbital and rotational dynamics on the climate of Kepler-62f using a 3-D GCM is therefore crucial for a more accurate assessment of its habitability.

	An understanding of the effect of tides on orbital evolution and planetary habitability is vital given the close proximity of low-mass planets orbiting in the habitable zones of their stars, and the likelihood of additional planetary companions in these systems. Tides raised between a star and a close orbiting planet introduce torques on the planet, resulting in changes in semi-major axis and eccentricity \citep{Barnes2008, Barnes2009}. With an orbital period of $\sim$267 days, tidal effects are expected to be weak, but could affect the rotation rate of the planet \citep{Heller2011}. Recent work has shown that planets orbiting in the habitable zones of lower-mass stars may not necessarily be synchronously rotating \citep{Leconte2015}. Bolmont \emph{et al.} (\citeyear{Bolmont2014, Bolmont2015}) found it likely that the rotation period of Kepler-62f is still evolving (and therefore not synchronized with its orbital period), and that it could have a high obliquity. However, they used only one tidal model, and others exist.

	Here we combine constraints obtained with an orbital evolution model with a 3-D GCM to explore the effect of orbital configuration on the climate and habitability of Kepler-62f, for different atmospheric compositions and planetary rotation rates. We calculated the orbital locations of the inner four planets in the Kepler-62 system relative to the position of Kepler-62f, for a range of possible initial eccentricities and longitudes of pericenter for this planet. We then used an \emph{n}-body model to integrate the orbits of all five planets given different planetary masses, to identify the maximum initial eccentricity possible for Kepler-62f while still maintaining stability within the planetary system. With these results we calculated the rotational and obliquity evolution possible for this planet. The potential habitability as a function of its stable eccentricity was then explored with a GCM, for atmospheres with Earth-like and high CO$_2$ levels, focusing on the presence of surface liquid water on the planet over an annual cycle.

\section{Models}
Here we describe the orbital and climate models used for this study. The existing \emph{n}-body model HNBody (Section 2.1) is used to integrate the orbits of multiple planets in a system orbiting a central, dominant body (in this case a star). Our method for estimating the masses for the planets is described in Section 2.2. In Section 2.3 we describe the models used to explore the tidal evolution of Kepler-62f. The two GCMs used to run climate simulations of the potentially habitable planet Kepler-62f (with output from the \emph{n}-body model as input) are described in Section 2.4.  

\subsection{HNBody: Modeling the Orbital Evolution of Multiple-Planet Systems}
	The Hierarchical N-Body (HNBody) package \citep{Rausch2002}, a set of software utilities we employed as our  \emph{n}-body model, integrates the orbits of astronomical bodies governed by a dominant central mass. It is based on the technique of symplectic integration, an \emph{n}-body mapping method, developed by Wisdom and Holman (\citeyear{Wisdom1991}), and performs standard point mass orbital integrations for a given number of planets in a system. 
	
	In the \emph{n}-body mapping method, the Hamiltonian\textemdash a mathematical function used to generate the equations of motion for a dynamical system\textemdash is the sum of the Keplerian and the interacting contributions to the motion of orbiting planetary bodies \citep{Wisdom1991}. The former describes the Keplerian motion of the orbiting planetary bodies around a central mass (the host star), and the latter describes the gravitational interactions between the planets themselves.

	The evolution of the complete Hamiltonian is determined by alternately evolving the Keplerian and interacting parts separately in a sequence of steps leading to new \emph{n}-body maps of the system, which are composed of the individual Keplerian evolution of the planets and the kicks due to the perturbations the planets deliver to one another. The output from HNBody consists of a series of data files that describe the evolution of selected orbital parameters over time. In the next section we describe the inputs that HNBody requires to generate this information.

\subsubsection{HNBody: Model Inputs}
	While we know from transit timing data where each planet in the Kepler-62 system is relative to our line of sight when it transits its host star, the locations of the other four planets at each planet's individual transit time are unconstrained. An accurate integration of their orbits requires the location of all planets at the same epoch. 
		
	The Keplerian orbit of a planetary body can generally be described by a set of six parameters that characterize the orbit. The parameters we chose as input to HNBody for each of the five planets in the Kepler-62 system were the semi-major axis $a$, the orbital eccentricity $e$, the inclination of the orbit $i$, the longitude of the ascending node $\Omega$, the longitude of pericenter $\omega$, and the true anomaly $f$ (see Figure 1 and Table 1 for definitions of these parameters). 
	
	While $a$ and $i$ are constrained from transit observations of the Kepler-62 system \citep{Borucki2013}, $e$, $\Omega$, $\omega$, and $f$ are poorly constrained. Given the close proximities of planets 62b-e (0.05-0.43 AU), which are all within their tidal circularization orbital radii after 4.5 Gyr \citep{Kasting1993}, and the estimated age of Kepler-62 ($\sim$7 billion years, \citealp{Borucki2013}), we assumed $e=0$ for these inner four planets. However, we acknowledge that in a multiple-planet system, the tidal evolution of a close-in planet is coupled to secular interactions with other planets in the system, and these interactions can cause planet eccentricities to vary \citep{Greenberg2011, Laskar2012, Hansen2015}. 
	
	Following from this assumption, $\omega$ is undefined, as all points along the orbit are equidistant from the star. We also assumed $\Omega=0$ for all planets. This is reasonable given that $i\sim$89-90$^\circ$ for all planets, thus constituting an edge-on orbit capable of yielding a transit observable by the Earth. We determined the locations of Kepler-62b-e relative to Kepler-62f at the same point in time, assuming a range of possible eccentricities between 0.0 and 0.9 and a range of longitudes of pericenter $\omega$ between $0$ and $2\pi$ for Kepler-62f. In the Appendix we outline the equations that were used to generate the locations of all planets in the system, using transit data for these planets \citep{Borucki2013}, and the aforementioned values for the other orbital elements. 	
	
	\subsection{Planetary masses of the Kepler-62 system}
	HNBody also requires a seventh parameter as input\textemdash the planet's mass relative to its host star. There are several mass-radius relations in the literature, and we chose two to explore the effect of planetary mass on the orbital evolution of the Kepler-62 system. We used the following mass-radius relation determined by Kopparapu  \emph{et al.} (\citeyear{Kopparapu2014}) derived from the exoplanets.org database \citep{Wright2011}:
	
\begin{equation}\label{mr}
	\frac{M_p}{M_\oplus} = 0.968 (\frac{R_p}{R_\oplus})^{3.2}, \hspace{10mm}	M_p<5M_\oplus
	\end{equation}
	
	This yields a planetary mass of $\sim$3M$_\oplus$ for Kepler-62f. We also ran HNBody integrations with planet masses derived using the following mass-radius (for Kepler-62d, e, and f) and density $\rho$ (for Kepler-62b and c) relations from Weiss \emph{et al.} (\citeyear{Weiss2014}):
	\begin{equation}\label{weissmass}
	\frac{M_p}{M_\oplus} = 2.69 (\frac{R_p}{R_\oplus})^{0.93}, \hspace{10mm} 1.5\leqslant \frac{R_p}{R_\oplus}<4 \hspace{10mm} \text{(gaseous)}
	\end{equation}
	
	\begin{equation}\label{weissdens}
	\rho = 2.43 + 3.39 (\frac{R_p}{R_\oplus})\hspace{3mm} g/cm^3, \hspace{10mm}R_p<1.5R_\oplus \hspace{10mm} \text{(rocky)}
	\end{equation}
	The Weiss \emph{et al.} (\citeyear{Weiss2014}) relationship resulted in higher masses for Kepler-62b and f, and lower masses for Kepler-62c, 62d, and 62e, compared to those using the Kopparapu  \emph{et al.} (\citeyear{Kopparapu2014}) treatment. Finally, we ran additional orbit integrations with the maximum mass limits for all planets, as determined by Borucki \emph{et al.} (\citeyear{Borucki2013}). Although these maximum masses are likely unphysical, they should result in a similar constraint on eccentricity as more physically plausible mass estimates (such as iron or Earth-like compositions). This varied approach allows us to address the spread in mass-radius relationships that can arise given different planetary compositions \citep{Wolfgang2015}. The masses for all five Kepler-62 planets used as input to HNBody integrations are given in Table 2. 

	Additional model specifications include the preferred coordinate system, the class of particles (based on the scale of the masses and how their interaction is to be taken into account), and the timestep and total length of integration. We specified a bodycentric coordinate system (ex. heliocentric reference frame in the case of the Solar System), which treats the system as dominated by the mass of the central star. We ran our HNBody integrations for interactions between ``heavy weight particles" (which includes the star and the planets) for $10^6$ years, with a time step equal to 1/20th and 1/100th of the orbital period of the innermost planet in the Kepler-62 system. As Kepler-62b has an orbital period of 5.7 days, we used time steps of 0.29 and 0.06 days, respectively. See Deitrick \emph{et al.} (\citeyear{Deitrick2015}) for more details on HNBody. Barnes and Quinn (\citeyear{Barnes2004}) showed that in simulations that last one million orbits, only $\sim$1\% of ejections occurred in the timescale between 10$^5$ and 10$^6$ years of simulation, with the vast majority of unstable configurations occurring within 10$^4$ years. While orbital instabilities can arise at any timescale, one million orbital periods has emerged as a practical reference. 
	
	We defined orbital stability as a successful integration in which stable, periodic amplitude oscillation was present for all planets throughout the entire million years, the energy was conserved to better than one part in 10$^4$, and no planets were ejected from the system. Eccentricities spanning the full range of stability for Kepler-62f were then used as input to our GCMs. We then ran climate simulations of this planet with a variety of atmospheric compositions, orbital configurations and rotation rates to explore and assess its possible climate states, and to determine the best possible combination of these parameters for surface habitability.

\subsection{Tidal Model}
In this section we consider the tidal evolution of Kepler-62f. Bolmont \emph{et al.} (\citeyear{Bolmont2014, Bolmont2015}) examined the rotational evolution of planets e and
f and found that the rotation period of f is not synchronized with the orbital period and that a wide range of obliquities is possible. However, they only considered circular orbits for planet f.
Furthermore, they used only one equilibrium tide model in which the time interval between the passage of the perturber and the passage of the tidal bulge is constant, a model we call the constant time lag (CTL) model \citep{Hut81,FerrazMello08,Leconte10}. In this section we relax some of these choices and find that synchronous rotation of Kepler-62f is possible.

In addition to the CTL model, we also employed a model in which the angle between the perturber and the tidal bulge, as measured from the center of the planet, is constant, a model we call the constant phase lag (CPL) model \citep{GoldreichSoter66,FerrazMello08,Heller2011}. These two equilibrium tide models are mathematical treatments of tidal deformation, angular momentum transfer, and energy dissipation due to tidal friction, and both have known flaws \citep{Greenberg09,EfroimskyMakarov13}. However, they provide a qualitative picture of tidal evolution and can be used to infer possible rotation states of Kepler-62f. 

Both models are one-dimensional with the tidal properties encapsulated in either the time lag $\tau$ for CTL or the tidal quality factor $Q$ for CPL. For both models we assumed that Kepler-62f has the same tidal properties as the Earth: $\tau_\oplus = 640$~s~\citep{Lambeck77} and $Q_\oplus = 12$~\citep{Yoder95}. The values for Kepler-62f are assuredly different, but these choices are likely not off by more than an order of magnitude, assuming Kepler-62f has oceans and continents. A $Q$ of 100 would result in a tidal locking timescale that is less than ten times longer. 

Rather than present the full set of equations, the reader is referred to Appendix E of \cite{Barnes2013}. The salient features are that each model contains 6 coupled differential equations that track the orbital semi-major axis and eccentricity, as well as the rotation rate and obliquity of both bodies. The effects of the other planets in the system are ignored. Therefore, we consider limiting cases to explore the range of these effects. The equations conserve angular momentum, and energy is dissipated by the tidal deformation of the bodies. In this case, tidal effects on the star are negligible.

We used calculated minimum and maximum initial eccentricities for Kepler-62f (Section 3); initial spin periods of 8 hr, 1 day, or 10 days; and initial obliquities of $5^\circ$, $23.5^\circ$, or $80^\circ$. We chose this range of initial conditions for Kepler-62f with the goal of setting boundary conditions for the climate simulations. 

 \subsection{Climate Modeling of Kepler-62f}
	  Our primary goal in the climate modeling of Kepler-62f was to identify the most favorable combination of planetary parameters that would result in areas of the planet with warm enough surface temperatures for liquid water. To determine the scenario for which the largest habitable surface area is possible, we varied the atmospheric composition, orbital eccentricity, planetary obliquity, the angle of the vernal equinox (the point in the planet's orbit where both hemispheres receive equal instellation) relative to pericenter (closest approach to the star), and the rotation rate of the planet in our GCM simulations. 
	  
 \subsubsection{Model Inputs to CCSM4}
	  
	  We used version 4 of the Community Climate System Model (CCSM4\protect\footnotemark{}), a fully-coupled, global climate model \citep{Gent2011}. We ran CCSM4 with a 50-meter deep, slab ocean (see e.g., \citealp{Bitz2012}), with the ocean heat transport set to zero, as done in experiments by Poulsen \emph{et al.} (\citeyear{Poulsen2001}) and Pierrehumbert \emph{et al.} (\citeyear{Pierrehumbert2011b}). The ocean is treated as static but fully mixed with depth. The horizontal resolution is 2$^\circ$. There is no land, hence we refer to it as an ``aqua planet". The sea ice component to CCSM4 is the Los Alamos sea ice model CICE version 4 \citep{Hunke2008}. We made the ice thermodynamic only (no sea-ice dynamics), and use the more easily manipulated sea-ice albedo parameterization from CCSM3, with the surface albedo divided into two bands, visible ($\lambda \leqslant$ 0.7 $\mu$m) and near-IR ($\lambda >$ 0.7 $\mu$m). We used the default near-IR and visible band albedos  (0.3 and 0.67 for cold bare ice and 0.68 and 0.8 for cold dry snow, respectively). For more details, see Shields \emph{et al.} (\citeyear{Shields2013}). \footnotetext{We used three components of CCSM4 in this work: CICE4, CAM4, and a slab ocean. For simplicity, we refer collectively to this suite of model components as CCSM4 throughout this work, as done in Shields \emph{et al.} (\citeyear{Shields2013,Shields2014}). Bitz \emph{et al.} (\citeyear{Bitz2012})\textemdash the first study to use the CCSM4 slab configuration\textemdash also used this convention.}
	  
	 Because CCSM does not allow a 267-day orbital period (the actual orbital period of Kepler-62f), the orbital period was set to 365 days, so the model would still simulate a full annual cycle. As atmospheric radiative and convective adjustment timescales are short compared to either orbital period, we do not expect this to make much difference in the overall climate. We used CCSM4 for continuity with our previous work, to evaluate the general climate behavior given changes in orbital parameters, and used LMD Generic GCM to simulate the climate of a planet with physical characteristics more closely like those of Kepler-62f. The details of the LMD Generic GCM are given in the following section. 
	  
	  Kepler-62f receives 41\% of the modern solar constant from its star, therefore significant amounts of CO$_2$ or other greenhouse gases may be required to keep temperatures above the freezing point of water on the surface, as is widely assumed for planets near the outer edge of their host stars' habitable zones \citep{Kasting1993}. An active carbon cycle capable of generating increased amounts of atmospheric CO$_2$ in response to decreasing surface temperatures \citep{Walker1981} would be a relatively straightforward means of maintaining habitable surface temperatures on the planet. More than $\sim$2 bars of CO$_{2}$ could accumulate in the atmosphere of a planet with an active carbon cycle before the maximum greenhouse limit for CO$_{2}$ is reached \citep{Pierrehumbert2010}. Large increases in atmospheric CO$_2$ concentration begin to have significant effects on convection, and the manner in which it adjusts the temperature lapse rate (the rate at which atmospheric temperature decreases with increasing altitude) on a planet. Additionally, CO$_2$ condensation becomes likely at levels of 1 to 2 bars, and collisional line broadening becomes important, increasing the infrared opacity of the atmosphere  \citep{Pierrehumbert2005}. These effects are neglected in Earth-oriented GCMs such as CCSM4. We therefore used the CCSM4 model to simulate only scenarios with an Earth-like atmospheric CO$_2$ concentration (400 ppmv), and allowed water vapor to vary throughout each simulation according to evaporation and precipitation processes. The rest of the atmospheric composition is preindustrial. We then used an additional GCM\textemdash the LMD Generic GCM\textemdash to simulate the climate of Kepler-62f with Earth-like as well as 1-12 bars of CO$_2$, as LMD Generic GCM contains parameterizations for addressing atmospheres with high CO$_2$ content.  
	  
	  It was important to consider the possibility that Kepler-62f may not have sufficient atmospheric CO$_2$ to keep surface temperatures above freezing. We therefore explored alternate means of creating habitable areas of the planet with lower, Earth-like CO$_2$ levels. Given the effects of planetary obliquity \citep{Ward1974, Williams1975} and eccentricity \citep{Berger1993, Berger2006} on seasonality and annual global instellation, the best possible scenario for habitability in the low-CO$_2$ case may be one in which Kepler-62f has a high obliquity and a high eccentricity. Additionally, an orbital configuration in which the hotter, summer months in a given hemisphere coincide with the pericenter of the planet's orbit could amplify the effects of high obliquity and eccentricity. To test this prediction, we ran simulations with CCSM4 (and LMD Generic GCM), assuming an aqua planet, with a range of different values for these parameters.
	  
	  We ran 30-year GCM simulations using CCSM4 with the input maximum initial eccentricity for stable HNBody integrations for Kepler-62f (see Section 3). We also ran additional simulations with lower eccentricity values to explore the effect of eccentricity on instellation and planetary surface temperature. We used a synthetic stellar spectrum from the Pickles Stellar Atlas \citep{Pickles1998}, with flux in the range of 1150-25000 Angstroms, and an effective photospheric temperature of 4887 K, which is close to the estimated effective temperature of Kepler-62 (4925 K, \citealp{Borucki2013}). The percentage of the total flux from the star was specified in each of the twelve incident wavelength bands in CAM4 (see Table 3), as done in Shields \emph{et al.} (\citeyear{Shields2013}, \citeyear{Shields2014}).  
	
	We also ran 40-yr CCSM4 simulations with an Earth-like obliquity of 23$^\circ$, and with an obliquity of 60$^\circ$. To explore the influence of the location of summer solstice relative to pericenter, we varied the angle of the vernal equinox relative to the longitude of pericenter (VEP), which governs the difference in instellation between southern hemisphere summer and northern hemisphere summer (Fig. 13). Because CCSM4 is parameterized for Earth-like conditions, we kept the radius, mass, and surface gravity of the planet equal to those of Earth in these simulations. 

\subsubsection{LMD Generic GCM}
We used the ``Laboratoire de M\'et\'eorologie Dynamique" (LMD) Generic GCM \citep{Hourdin2006}, developed to simulate a wide range of planetary atmospheres and climates. It has been used in studies of the early climates of solar system planets \citep{Charnay2013, Forget2013, Wordsworth2013}, and in previous studies of the climates of extrasolar planets \citep{Wordsworth2011, Leconte2013, Wordsworth2015}.

	Like CCSM4, LMD Generic GCM solves the primitive equations of fluid dynamics using a finite difference method and a 3-D dynamical core. The radiative transfer scheme is based on a correlated-\textit{k} method, with absorption coefficients calculated from high-resolution spectra generated using the HITRAN 2008 database \citep{Rothman2009}. A smaller database of correlated-\textit{k} coefficients was then generated from this spectra using 12$\times$9$\times$8 temperature (T=100, 150, ...600, 650 K, in 50 K steps), log-pressure (p=0.1, 1, 10, ..., 10$^7$ Pa), and water vapor mixing ratio (q$_{H_2O}$=10$^{-7}$, 10$^{-6}$, ..., 1) grids. These correlated-\textit{k} coefficients ensure expedient radiative transfer calculations in the GCM. The spectral intervals included 38 shortwave (incident stellar radiation) bands and 36 longwave (outgoing radiation) bands, with a sixteen-point cumulative distribution function for integration of absorption data within each band. The radiative transfer equation is then solved in each atmospheric layer using a two-stream approximation \citep{Toon1989}. Parameterizations for convective adjustment, and CO$_2$ collision-induced absorption are included based on work by Wordsworth \emph{et al.} (\citeyear{Wordsworth2010a}). Simulations were run with a two-layer ocean, including a 50m-deep top layer and an underlying 150m-deep second layer. Both layers are assumed to be well-mixed, with horizontal diffusion used to approximate ocean heat transport by large-scale eddies. Adiabatic adjustment is also included, whereby the ocean lapse rate is adjusted to maintain a warmer top ocean layer at all times. 

	We ran our LMD Generic GCM simulations for 60-200 years (depending on the amount of CO$_2$ and required model equilibration time) with a horizontal spatial resolution of 64$\times$48 (corresponding to 5.625$^\circ$ longitude $\times$ 3.75$^\circ$ latitude), with 20 vertical levels. A blackbody spectrum with the stellar effective temperature of Kepler-62 (4925 K, \citealp{Borucki2013}) was used as the host star spectrum. The albedo of snow was set to 0.55 for the high-CO$_2$ simulations, and 0.43 for the simulations with Earth-like CO$_2$ (to match the two-band albedo parameterization in CCSM4 for accurate comparison). The albedo of sea ice was allowed to vary between a minimum of 0.20 and a maximum of 0.65, depending on ice thickness. We assumed an aqua planet with different amounts of atmospheric CO$_2$, and water vapor that varied throughout each simulation as with CCSM4. We used the radius of Kepler-62f (1.41 R$_\oplus$, \citealp{Borucki2013}), a surface gravity of 14.3 $m/s^2$, based on a  $\sim$3M$_\oplus$ planet (see Section 2.2), a 267-day year, Earth-like (24-hr) and synchronous (267-day) rotation rates, and the minimum and maximum stable initial eccentricities possible for Kepler-62f (see Section 3). Additional input parameters for simulations using CCSM4 and LMD Generic GCM for our work on Kepler-62f are given in Tables 4 and 5. The results of these simulations, and the implications for the habitability of Kepler-62f are presented and discussed in the following section. 

\section{Results} 
\subsection{N-body simulations}
	Figure 2 shows the fraction of stable orbital integrations performed using HNBody, assuming different initial eccentricities for Kepler-62f, and zero eccentricity for Kepler-62b-e (see e.g., \citealp{Barnes2001}). The maximum initial eccentricity for which stable integrations were possible for greater than 90\% of the simulated longitudes of pericenter, was $e=0.32$, assuming the Kopparapu \emph{et al.} (\citeyear{Kopparapu2014}) mass-radius relation. A higher upper eccentricity limit is 0.36 assuming a smaller mass (equal to that of the Earth) for Kepler-62f and a larger mass for Kepler-62e. Simulations using a shorter time step of 1/100th of the orbital period of the innermost planet reduced the percentage of longitudes of pericenter with stable integrations at $e=0.32$ by $\sim$30\%, though we still found over half of our simulated configurations at this eccentricity to be stable. Stable integrations were possible for 23\% of the simulated longitudes of pericenter at $e=0.33$. However, we did not consider this a large enough fraction of stable configurations, and consider $e=0.32$ to be the conservative maximum. The maximum stable initial eccentricity decreased by $\sim$3\%, to $e=0.31$, when the Weiss \emph{et al.} (\citeyear{Weiss2014}) mass-radius and density relations were used, and decreased  by 28\%, to $e=0.23$, using the maximum mass limits for all planets \citep{Borucki2013}. The evolution of the eccentricities of all five planets for a stable integration at $e=0.32$ for Kepler-62f is shown in Figure 3. All planets exhibited eccentricity oscillations\textemdash a characteristic of multiple-planet systems\textemdash with a regular period that remained constant over the entire $10^6$-year integration. Integrations assuming higher initial eccentricities for Kepler-62f yielded eccentricity evolution that exhibited irregular oscillatory motion indicative of the occurrence of significant orbital shifts. Consequently, we considered $0.00 \leqslant e \leqslant 0.32$ to be the maximum allowable range of initial eccentricities for Kepler-62f. 
	
\subsection{Tidal Evolution of Kepler-62f}
We began our simulations with $e = 0.00$ or 0.32, and spin periods and obliquities as given in Section 2.3. In general tidal evolution is faster for larger eccentricity, larger spin period, and smaller obliquity. Thus the $e=0.32$, 10 day spin period and $5^\circ$ obliquity cases should evolve most rapidly toward an equilibrium spin state.

In Figure 4 we show the evolution of spin period and obliquity for the two tidal models. The difference between the two models naturally produces different torques on the planetary rotation, and thus different timescales to reach equilibrium. On the left, the CTL model shows results very similar to those in \cite{Bolmont2014, Bolmont2015} and it is unlikely that the system has reached a tidally locked state, though tidal de-spinning is significant over 10 Gyr. The right panels show that tidal locking is possible, which is easily shown by the spin period evolution flattening out at either 267 days or 178 days. The latter corresponds to a 2:3 spin-orbit resonance which is closer to the equilibrium spin period for an eccentricity of 0.32. Only the extremely fast rotating and high obliquity cases do not tidally lock within 5 Gyr in the CPL model.

As the estimated age of Kepler-62 (7 Gyr, \citealp{Borucki2013}) is uncertain, as well as the initial eccentricity and rotation state, we conclude that the rotational period of Kepler-62f can lie anywhere from less than 1 day all the way to 267 days. Similarly, the obliquity can have a range of values from 0 to at least $90^\circ$. For our assumptions, the rotation period and
obliquity should lie betwen the solid gray line and the dotted black line. Therefore, habitability and climate studies of this planet should account for this range of rotational states, as well as both small and large obliquities. We present the results of such studies in the following section.

\subsection{Climate simulations}
	With the present atmospheric level (PAL) of CO$_2$ on Earth (400 ppmv), all CCSM4 simulations with eccentricity $e=0.00-0.32$ resulted in completely ice-covered conditions, with global mean surface temperatures below 190 K. While our CCSM4 simulations were run with a 365-day orbital period rather than the actual orbital period of Kepler-62f (267 days), a comparison of LMD Generic GCM sensitivity tests run with 267- and 365-day orbital periods showed equivalent results. Figure 5 shows the annual mean instellation as a function of latitude for these different eccentricities, assuming an obliquity of 23$^\circ$. The average instellation over an annual cycle increases with eccentricity \citep{Berger1993, Williams2002, Berger2006}, in accordance with the following relation:
\begin{equation}\label{ecc}
S = \frac{S_a}{\sqrt{1-e^2}}
\end{equation}
where $S$ is the average instellation at the mean star-planet distance; $S_a$ is the instellation at a given distance $a$ from the star during a planet's orbit; and $e$ is the eccentricity of the planet's orbit \citep{Berger2006}. 	
	
	If CO$_2$ is efficiently outgassed, and the silicate weathering rate is weak, higher levels of CO$_2$ may be expected to accumulate in the planet's atmosphere prior to reaching the maximum greenhouse limit \citep{Kopparapu2013b, Kopparapu2013a}. Since PAL CO$_2$ was clearly insufficient to yield habitable surface temperatures for Kepler-62f at its value of incoming stellar flux, we ran simulations using LMD Generic GCM with CO$_2$ concentrations of up to 12 bars, at both limits of the stable eccentricity range for Kepler-62f. We let the amount of water vapor vary (through transport, evaporation, and circulation) during the course of each simulation. Global mean, minimum, and maximum surface temperatures for each simulation are plotted in Figure 6. With 5 bars of CO$_2$ in the atmosphere, Kepler-62f exhibits a global mean surface temperature similar to modern Earth, at $\sim$282 and $\sim$290 K for the lower and upper stable eccentricity limits, respectively. Surface temperatures increase with CO$_2$ concentration. However, while the surface temperature on the planet is 46-50 K warmer (depending on eccentricity) with 5 bars of atmospheric CO$_2$ compared to the 3-bar CO$_2$ case, it is only an additional 20-25 K warmer ($\sim$308 K) with 8 bars of CO$_2$ in the atmosphere. Further increases in atmospheric CO$_2$ yield successively smaller increases in surface temperature. We also found that the planetary albedo reached a minimum at 8 bars, and then began to slowly increase (Figure 7). This indicates that we are likely approaching the point where the effects of Rayleigh scattering begin to dominate over the greenhouse effect at this level of CO$_2$. Additionally, the difference between minimum and maximum surface temperatures decreases with CO$_2$ concentration, highlighting the role of a denser atmosphere in evening out temperature contrasts \citep{Wordsworth2011, Pierrehumbert2011a}.

	While the majority of our simulations run with 3 bars of CO$_2$, including that run at Earth's present obliquity yielded globally ice-covered conditions (Figure 8), we did find that at an extremely high planetary obliquity (90$^\circ$), at the upper eccentricity limit (0.32), surface temperatures exceeded the freezing point of water over close to 20\% of the planet, though the global mean surface temperature was well below freezing, at 246 K. As shown in Figure 8, the regions with zero ice cover occur in the polar regions, which receive more instellation than the equator at high obliquities. We also found that, while high-eccentricity simulations with an obliquity of 60$^\circ$ ultimately yielded ice-covered conditions, maximum surface temperatures reached 273 K during the planet's orbit at small orbital distances from its star, which could result in melting of ice formed when the planet is at more distant points in its orbit.

	Figure 9 shows the surface temperature as a function of latitude and longitude for an LMD Generic GCM simulation with 5 bars of CO$_2$, variable water vapor, and $e=0.00$. With a global mean surface temperature of 282 K, only $\sim$17\% of the planet is ice-covered. Simulations run at $e=0.32$ resulted in a higher global mean surface temperature by $\sim$8 K, and $\sim$10\% less ice cover on the planet.   
	
	Given that Kepler-62f could have a wide range of rotation periods, including a synchronous rotation rate, using CCSM4 and LMD Generic GCM we ran simulations of the planet assuming an Earth-like rotation rate, and a synchronous rotation period, for 400 ppmv (Earth-like), 1-bar, and 3-bar CO$_2$ atmospheres, with both the eccentricity and obliquity set to zero. Our CCSM4 simulations with Earth-like CO$_2$ levels were completely ice covered. Figure 10 shows the surface temperature as a function of latitude and longitude for the LMD Generic GCM 1-bar and 3-bar CO$_2$ cases. The synchronous case with 1-bar of CO$_2$ has a global mean surface temperature of $\sim$206 K, with 99.7\% of the planet covered in ice. There is a small circular region of open water at the substellar point on the planet (Fig. 10). However, as we have not included sea ice transport in our GCM, and glacial flow of thick sea ice could cause a planet to become fully glaciated \citep{Abbot2011}, this region of open water is likely unstable. Furthermore, temperatures on the night side of the planet reach below the limit for CO$_2$ to condense at 1-bar surface pressure, indicating the likelihood for atmospheric freeze-out at this atmospheric concentration. The non-synchronous case with the same CO$_2$ level has a global mean surface temperature that is $\sim$11 degrees warmer ($\sim$217 K), although it is completely covered in ice. The 3-bar CO$_2$ cases, while $\sim$17-18 K warmer, exhibit a similar behavior, with the non-synchronous case yielding a global mean surface temperature ($\sim$235 K) that is $\sim$12 K warmer than the synchronous case ($\sim$223 K). Surface temperatures on the night side of the planet in the 3-bar CO$_2$ case are right at the boundary for CO$_2$ condensation at this surface pressure and CO$_2$ concentration.	

	High amounts of CO$_2$ would be a relatively straightforward means of generating habitable surface temperatures on a planet receiving low instellation from its star. However, the efficiency of the carbonate-silicate cycle\textemdash which has been shown to be sensitive to a variety of factors, including the mantle degassing rate of a planet \citep{Driscoll2013}\textemdash is unknown for Kepler-62f. If the planet has a low amount of CO$_2$ in its atmosphere, and lacks an active carbon cycle to adjust the silicate weathering rate with temperature \citep{Walker1981}, the fraction of habitable surface area may decrease significantly compared to our simulated cases with 3 bars and higher CO$_2$. We therefore explored different orbital configurations which could improve conditions for habitability on a planet that does not have an effective means of increasing its concentration of greenhouse gases. 
	
	Since we do not know the location of pericenter for the orbit of Kepler-62f, we explored the effect on instellation of changes in the VEP of a planet's orbit, as this can affect the hemispherical annually-averaged instellation on a planet \citep{Berger1978, Berger1993}. Figure 11 shows the results of CCSM4 simulations assuming a VEP of 0$^\circ$ and 90$^\circ$, with the obliquity and eccentricity held fixed at 23$^\circ$ and 0.32, respectively. At VEP=0$^\circ$, the point where both hemispheres receive equal amounts of instellation coincides with the planet's closest approach to its star. The difference in monthly instellation is relatively small in the southern hemisphere, as it receives equal instellation to that received by the northern hemisphere at pericenter. At a VEP of 90$^\circ$, the southern hemisphere receives a significantly higher percentage of instellation compared to the northern hemisphere during its summer months when the planet is at or near pericenter. Because of the planet's obliquity, this is when the southern hemisphere is angled toward the star. 

	The obliquity of Kepler-62f is observationally unconstrained, so we also explored how different obliquities might affect the planet's climate in a low-CO$_2$ scenario. As shown in Figure 12, CCSM4 at an obliquity of 60$^\circ$ reveals more instellation received in the high-latitude regions of the planet than in the tropics. The global mean surface temperature is still significantly below freezing at both of the simulated obliquities (23$^\circ$ and 60$^\circ$), given the Earth-like CO$_2$ levels and low stellar flux. However, surface temperatures do get above freezing in the southern hemisphere during its summer months in the high-obliquity case, with VEP=90$^\circ$, in both our CCSM4 and LMD Generic GCM simulations. This means that summer in the southern hemisphere occurs near the planet's closest approach to its star, as shown in the schematic diagram in Figure 13. This results in higher annual mean surface temperatures in this hemisphere during its summer months, compared to the northern hemisphere. The less extreme cold temperatures in the LMD Generic GCM simulations are due to a 10-meter maximum thickness limit of sea ice in LMD Generic GCM, while the CCSM4 sea ice thickness near the poles is $\sim$30 meters. This causes a greater temperature difference between the models at the poles during winter months. The reduced ice thickness in LMD Generic GCM results in a larger conductive heat flux through the ice, though this has a lower impact for the warmer climates (e.g. with 3-5 bars and higher CO$_2$), where there is less ice. We calculated the temperature difference between the models by equating the residual surface energy flux with the conductive heat flux through the ice assuming 10- and a 30-m ice thicknesses, and found the difference to equal that between the coldest temperatures in LMD Generic GCM and CCSM4 ($\sim$60 K). Both models indicate that an orbital configuration that places the summer solstice near pericenter may amplify the effects of high obliquity and eccentricity, and this may cause surface melting to occur during an annual cycle in a low-CO$_2$ scenario. This effect is also seen clearly in a comparison between LMD Generic GCM simulations with 3 bars of atmospheric CO$_2$ and VEP of 0$^\circ$ and 90$^\circ$ (Figure 14), where the latter simulation shows warmer surface temperatures in the southern hemisphere of the planet. We noticed a decrease in planetary ice cover of $\sim$0.6\% per 90-degree increase in VEP in our 3-bar CO$_2$ simulations run at 90$^\circ$ obliquity and $e=0.32$.   

\section{Discussion}
	We explored the plausible range of orbital, rotational, and atmospheric states of Kepler-62f and found that some permit habitability, but some do not. A global mean surface temperature similar to modern-day Earth is possible throughout the planet's orbit with 5 bars of CO$_2$ in its atmosphere, for both lower and upper limits of the range of stable initial eccentricities ($e=0.00$ and $e=0.32$), assuming present Earth obliquity. Our simulations with $e=0$ yielded lower global mean surface temperatures, consistent with average instellation decreasing with decreasing eccentricity. These results indicate that if Kepler-62f has an active carbon cycle, where CO$_2$ is allowed to build up in the atmosphere as silicate weathering decreases at lower surface temperatures \citep{Walker1981}, the best possible scenario for habitable surface conditions is one that combines moderate to high eccentricity with high atmospheric CO$_2$. The maximum CO$_2$ greenhouse limit for stars with the effective temperature of Kepler-62 occurs at a stellar flux\protect\footnotemark{} that is well below that received by Kepler-62f (41\% of the modern solar constant, \citealp{Borucki2013}). Given that 3-5 bars of CO$_2$ are significantly below this limit  ($\sim$7-8 bars, \citealp{Kopparapu2013b, Kopparapu2013a}), these levels of CO$_2$ are plausible for this planet. 
\footnotetext{$\sim$30\% of the modern solar constant \citep{Kopparapu2013b, Kopparapu2013a}}	

	Earlier work exploring the possible climates of Kepler-62f proposed habitable conditions with 1.6 bars of CO$_2$ or more in the planet's atmosphere \citep{Kaltenegger2013}. However, this work was not done with a 3-D GCM, and therefore lacked a full treatment of clouds and atmosphere-ocean interactions. Using a 3-D GCM, we find that the only scenario that permits Kepler-62f to exhibit clement temperatures for surface liquid water with 3 bars of CO$_2$ in the planet's atmosphere requires stringent orbital configuration requirements (high eccentricity and obliquity). With an obliquity of 90$^\circ$, open water was present over just $\sim$20\% of the planet, at polar latitudes. This indicates that it is likely that more than 1.6 bars of CO$_2$ is required for surface habitability on this planet.
	
	We assumed a five-planet system for Kepler-62 in our \emph{n}-body simulations. While there is currently no observational evidence for additional planets, the existence of other planets in this system is a clear possibility, and their presence would certainly affect the eccentricity limit for dynamical stability that we have calculated here. Future discovery of additional planets in this system with transit timing variations or RV may provide additional dynamical constraints on the eccentricity of Kepler-62f. 
	
	We combined climate simulations using CCSM4 and the LMD Generic GCM to provide a comprehensive exploration of the possible climates for Kepler-62f given its \emph{n}-body model constraints. Previous work using the LMD Generic GCM at a higher resolution (e.g. 128$\times$96) yielded similar results to those at the resolution we employed here (64$\times$48), and as we confirmed that the global mean surface temperature had not changed by more than 1$^\circ$K in the last 20 years of simulation, we are confident that running our simulations for longer timescales would not change our results significantly. We verified that simulations assuming Earth-like CO$_2$ levels in both GCMs exhibited similar (ice-covered) conditions, confirming the robustness of these simulations to various assumptions. 
	
	Our simulations with higher levels of CO$_2$ resulted in increasingly higher surface temperatures on the planet. As CO$_2$ was increased, the changes in global mean surface temperatures became progressively smaller. This logarithmic relationship between CO$_2$ concentration and radiative forcing is long established \citep{Wigley1987}. As regions of the spectrum become opaque, additional CO$_2$  molecules become far less effective at increasing temperatures \citep{IPCC1990}. Thus, the greenhouse effect starts to become less efficient as a warming mechanism. Additionally, CO$_2$ is $2.5\times$ more effective as a Rayleigh scatterer than Earth's air \citep{Kasting1991, Forget1997}, and this behavior likely contributes to the loss of warming at higher CO$_2$ concentrations \citep{Kasting1991, Kasting1993, Selsis2007}. 		
	
	We have not included the effect of CO$_2$ condensation in our simulations. As discussed in Section 2.4, at levels of 1-2 bars, CO$_2$ condensation is likely to occur in the upper atmosphere \citep{Pierrehumbert2005}. Depending on the particle size of CO$_2$ ice grains, this could result in cooling of the planet due to the albedo effect of CO$_2$ ice clouds \citep{Kasting1991}, or warming by scattering outgoing thermal radiation back towards the surface of the planet \citep{Forget1997}. 

	Kopparapu \emph{et al.} (\citeyear{Kopparapu2013b, Kopparapu2013a}) found that the maximum CO$_2$ greenhouse limit is $\sim$7-8 bars for a star with a similar effective temperature to that of Kepler-62 \citep{Kopparapu2013b, Kopparapu2013a}. We did find that the planetary albedo, which had decreased with increasing CO$_2$ concentration, started to increase, albeit slowly, at a CO$_2$ level of 8 bars. However, our simulations with 8-12 bars of CO$_2$ resulted in global mean surface temperatures that were still higher than those with lower CO$_2$, so we conclude that we had not yet reached the maximum CO$_2$ limit in our simulations, and that it may be higher than originally proposed. Kopparapu \emph{et al.} (\citeyear{Kopparapu2013b, Kopparapu2013a}) used a 1-D radiative-convective model in their work, and did not include the effect of water clouds or CO$_2$ clouds in their calculations.  While water clouds could increase the planetary albedo, thereby cooling the planet further, they may also contribute to the greenhouse effect, as both H$_2$O and CO$_2$ have strong absorption coefficients in the near-IR, which increase the amount of radiation absorbed by planets with lower-mass host stars that emit strongly in the near-IR \citep{Kasting1993, Selsis2007, Kopparapu2013b, Kopparapu2013a, Joshi2012, Shields2013, Shields2014}. Our results with a 3-D GCM do include water clouds, though not CO$_2$ clouds. A comprehensive study of the effect of CO$_2$ condensation as a function of particle size would be an important step towards identifying its ultimate effect on planetary climate. Regardless, dense CO$_2$ atmospheres can be expected to cause more even distributions of heat across a planet, and reduce the contrast between maximum and minimum surface temperatures. This is particularly important on synchronously-rotating planets \citep{Joshi1997, Edson2011}, where the difference in instellation on the day and night sides of the planet is large. We also see this result in our simulations, which show smaller maximum/minimum surface temperature contrasts with larger amounts of  CO$_2$. 
	
	All of our simulations assumed an aqua planet configuration, with no land. Previous work exploring the habitability of planets composed almost entirely of land and orbiting G-dwarf stars suggests that due to their lower thermal inertia and drier atmospheres, land planets are less susceptible to snowball episodes than aqua planets, requiring 13\% less instellation to freeze over entirely \citep{Abe2011}. The presence of land could certainly affect the silicate weathering rate and atmospheric concentration of CO$_2$ on a planet. However, Abbot \emph{et al.} ({\citeyear{Abbot2012}) found that climate weathering feedback does not have a strong dependence on land fraction, as long as the land fraction is at least 0.01. Edson \emph{et al.} (\citeyear{Edson2012}) found that the amount of CO$_2$ that accumulates in the atmosphere of a synchronously-rotating planet could be much greater if the substellar point is located over an ocean-covered area of the planet, where continental weathering is minimal, although atmospheric CO$_2$ concentration could still be limited by seafloor weathering processes \citep{Edson2012}. Including land in future simulations of a synchronously-rotating Kepler-62f, using a GCM with a carbonate-silicate cycle included (rather than assigning a prescribed atmospheric CO$_2$  concentration as we have done here) would be a valuable step towards assessing the role of surface type in regulating the atmospheric CO$_2$ inventory on synchronously-rotating planets. 
	
	We have concentrated on CO$_2$ as the primary greenhouse gas in our study of the influence of atmospheric composition on the habitability of Kepler-62f. Previous work has highlighted the role of molecular hydrogen as an incondensable greenhouse gas that, due to collision-induced absorption, could allow planets to maintain clement temperatures for surface liquid water far beyond the traditional outer edge of a star's habitable zone \citep{Pierrehumbert2011c}. Given that the orbital distance of Kepler-62f is interior to the limit for which an entire primordial H$_2$ envelope could be lost due to extreme ultraviolet-driven atmospheric escape (assuming a 3-4 M$_\oplus$ planet), even at apocenter (at an eccentricity of 0.32), an H$_2$ greenhouse may not be an effective mechanism for warming this planet. Exploring this mechanism in detail in the context of the Kepler-62 system would be an interesting topic for future study.
	
\subsection{Additional orbital influences on the climate of Kepler-62f}
	High amounts of CO$_2$ in the atmosphere of Kepler-62f would provide for consistently habitable surface conditions throughout the planet's orbit at its level of instellation, thereby offering the best chances for sustained surface liquid water and life. However, given the uncertainty of an active carbonate-silicate cycle operating on this planet, we explored orbital parameters that may periodically improve surface conditions during the course of an orbit in the event of low-CO$_2$ atmospheric conditions similar to the Earth, though these conditions would be short-lived. Planetary habitability throughout the course of an eccentric orbit has been shown to be most strongly affected by the time-averaged global instellation\textemdash provided there is an ocean to contribute to the planet's heat capacity\textemdash which is greater at higher eccentricities \citep{Williams2002}. Therefore, although high-eccentricity planets may spend significant fractions of an orbit outside of their stars' habitable zones, high eccentricity may help these planets maintain habitable surface conditions over an annual cycle \citep{Kopparapu2013b, Kopparapu2013a}, though habitability could be affected by tidal forces and resultant heating at close distances from the star during a planet's eccentric orbit \citep{Driscoll2015}. Planets on eccentric orbits could undergo freeze/thaw cycles depending on the orbital distance of apocenter and pericenter, respectively, and this could also have important implications for planetary habitability. Limbach and Turner (\citeyear{Limbach2015}) found that catalogued RV systems of higher multiplicity exhibit lower eccentricities. However, as these are statistical results, they do not mean that all such systems have low eccentricities. Furthermore, they used data from RV systems of more massive planets than those in the Kepler-62 system. There are currently no observational constraints on the eccentricities of low-mass, long-period extrasolar planetary systems. Our calculated eccentricity of 0.32 is entirely plausible for Kepler-62f, and consistent with observations. 
	
	Given that the obliquity, rotation rate, and the VEP for Kepler-62f are unknown, we explored how these factors might also influence surface habitability on this planet. Our results indicate that high obliquity, which has been shown to increase seasonality and stability against snowball episodes \citep{Williams1997, Williams2003, Spiegel2009}, results in even higher seasonality at high eccentricity, due to the larger difference between the orbital distance at pericenter and apocenter \citep{Williams2002}. This effect is more pronounced in the southern hemisphere during its summer months, when the angle of the vernal equinox relative to pericenter is 90$^\circ$, so that the planet's high obliquity significantly increases the instellation received by this hemisphere. If we had simulated VEP=270$^\circ$ we would expect the northern hemisphere to exhibit a similar effect, as this is a symmetric problem. Though we did not include ocean heat transport in our high-obliquity CCSM4 simulations that yielded global ice cover, previous work has found that snowball collapse is possible at high obliquity regardless of the presence of a dynamical ocean with ocean heat transport \citep{Ferreira2014}. 
	
	We identified several orbital configurations that, though rare, may cause temporarily habitable surface conditions during the course of an orbit given Earth-like levels of CO$_2$. Even on an ice-covered high-obliquity planet, our simulations using both models showed that surface temperatures reached above the freezing point of water during southern hemisphere summer months, which could cause surface melting. Our LMD Generic GCM simulations with 3 bars of CO$_2$ also resulted in higher surface temperatures in the summer hemisphere with a higher VEP of 90$^\circ$ compared to 0$^\circ$. The VEP can therefore amplify the warming effects of high obliquity and eccentricity, and may keep ice from forming at the poles, or reducing an ice sheet formed during the planet's orbit. If the planet experiences large oscillations in obliquity, polar ice could be prevented from forming on both hemispheres over an annual cycle, allowing habitable surface conditions to be maintained on planets with large eccentricities \citep{Armstrong2014}. 
	
	The weaker Coriolis force on synchronously-rotating planets permits rapid heat transport by advection compared to radiative heat transfer for surface pressures significantly greater than $\sim$0.2 bars \citep{Showman2013}. Combined with sufficient greenhouse gas concentration levels \citep{Joshi1997}, this can prevent atmospheric freeze out on the night side of the planet, through the transport of high amounts of heat from the day side to the night side \citep{Pierrehumbert2011a}. Our results have shown that a synchronously rotating  Kepler-62f exhibits a lower global mean surface temperature than one that is non-synchronous. This may underscore a lack of sufficient heat distribution between the day and night sides of the planet at the levels of CO$_2$ which we have simulated at this planet's instellation. Horizontal heat transport is less effective for smaller amounts of CO$_2$ \citep{Wordsworth2011}, which we confirmed in the 1-bar CO$_2$ case, where the ratio between the outgoing thermal fluxes on the night side and day side (a proxy for the redistribution efficiency) is lower. This caused larger differences between maximum and minimum surface temperatures on the day and night sides of the 1-bar CO$_2$ planet compared to the 3-bar CO$_2$ case. 
	
	Additionally, though we did not measure significant differences in planetary albedo between the synchronous and non-synchronous climate simulations at a given CO$_2$ concentration, Kepler-62f lies near the outer edge of its star's habitable zone, receiving 41\% instellation. On more temperate synchronously-rotating planets than we have simulated here, the increased cloud cover that will likely result on the day side of a synchronously-rotating planet with an ocean could increase the overall planetary albedo and reduce surface temperatures \citep{Yang2013,Yang2014}. The strength of this effect on distant worlds would depend on the greenhouse gas concentration and the behavior of the hydrological cycle. The results presented here imply that synchronously-rotating planets may require more CO$_2$ in their atmospheres than their non-synchronous counterparts to generate equivalent global mean surface temperatures farther out in their star's habitable zones, and this could affect planetary habitability.
	
	The \emph{n}-body model used in this work does not include the effect of tides. With an orbital period of 267 days, tides are likely to be weak on Kepler-62f, but they could affect its rotation period (see Section 3.2), and would certainly be an important consideration for potentially habitable planets orbiting even closer to their stars. Tidal effects can lead to changes in orbital parameters, and may induce capture into resonances in spin-orbit period, depending on the planet's eccentricity and its equatorial ellipticity (the equilibrium shape attained as a result of the gravitational interaction between the planet and the host star, \citealp{Rodriguez2012}). Our work here explored the extreme case of spin-orbit resonance\textemdash synchronous rotation. Non-synchronous resonance configurations, such as the 3:2 spin-orbit resonance observed on the planet Mercury \citep{Goldreich1966, Correia2004} are also possible, and may be sustained on a planet in a non-circular orbit over long timescales \citep{Rodriguez2012}. Such configurations become more likely at larger orbital eccentricities \citep{Malhotra1998, Correia2004}, and are worth exploring.
	
	The habitability requirements we have determined here did not account for any uncertainties in measured parameters such as stellar luminosity and semi-major axis. Including such uncertainties might permit a larger habitable surface area for an even wider range of parameters.

	Using constraints from \emph{n}-body models as input to GCM simulations permits the exploration of the impact on climate of the gravitational interactions inherent to a growing population of exoplanets\textemdash potentially habitable planets orbiting in multiple-planet systems around low-mass stars\textemdash and the resulting prospects for the habitability of these worlds. The techniques presented here can be applied to planets orbiting stars of any spectral type, with a range of possible atmospheric and surface compositions and dynamical architectures. They can be used to help assess the potential habitability of newly discovered planets for which observational measurements are still limited, and can be easily modified to incorporate new observational data that are acquired for these planets in the future. While future missions such as the Transiting Exoplanet Survey Satellite ({\it TESS}, \citealp{Ricker2009, Ricker2014}) and the James Webb Space Telescope \citep{Gardner2006} will not be capable of characterizing distant planets like Kepler-62f, the procedure we have carried out here serves as excellent preparation for closer planets that could be observed and characterized with these missions. These methods will help identify which among the habitable-zone planets discovered to date could exhibit conditions conducive to the presence of surface liquid water for the widest range of atmospheric and orbital conditions, making them priorities for these and other future characterization missions.
	
\section{Conclusions}
	We carried out a comprehensive exploration of the orbital evolution of Kepler-62f using an \emph{n}-body model, and found that the range of eccentricities that Kepler-62f could have while maintaining dynamical stability within the system was $0.00 \leqslant e \leqslant 0.32$. The upper limit of 0.32 is consistent with the analytic Hill Stability criterion
(cf. \citealp{Gladman1993, Barnes2006}). A higher upper eccentricity limit is 0.36 assuming a smaller mass (equal to that of the Earth) for Kepler-62f and a larger mass for Kepler-62e. The constraints from the \emph{n}-body model were used as input to 3-D climate simulations to explore the possible climates and habitability of Kepler-62f. 
	
	At 41\% of the modern solar constant, this planet will likely require an active carbonate-silicate cycle (or some other means by which to produce high greenhouse gas concentrations) to maintain clement conditions for surface liquid water. With 3 bars of CO$_2$ in its atmosphere and an Earth-like rotation rate, 3-D climate simulations of Kepler-62f yielded open water across $\sim$20\% of the planetary surface at the upper limit of the stable eccentricity range possible for the planet, provided that it has an extreme obliquity (90$^\circ$). With 5 bars of CO$_2$ in its atmosphere, a global mean surface temperature similar to modern-day Earth is possible for the full range of stable eccentricities and at the present obliquity of the Earth. This higher CO$_2$ level is therefore optimal, as it is below the maximum CO$_2$ greenhouse limit, and generates habitable surface conditions for a wide range of orbital configurations throughout the entire orbital period of the planet. If Kepler-62f is synchronously rotating, CO$_2$ concentrations above 3 bars would be required to distribute sufficient heat to the night side of the planet to avoid atmospheric freeze-out. 
			
	We have also shown that surface temperatures above the freezing point of water during an annual cycle are possible on the planet if it has a low (Earth-like) level of CO$_2$, provided that the obliquity is high (60$^\circ$ or greater) compared to an Earth-like obliquity (23$^\circ$), and the summer solstice at a given hemisphere occurs at or near the planet's closest approach to its star. This is a rare but possible orbital configuration that could cause surface melting of an ice sheet formed during a planet's orbit, and amplify the effects of moderate to high eccentricity and obliquity. While less optimal than the high-CO$_2$ case, this may result in periodically habitable surface conditions in a low-CO$_2$ scenario for Kepler-62f.

\section{Acknowledgments}

This material is based upon work supported by the National Science Foundation under Award No. 1401554, and Grant Nos. DGE-0718124 and DGE-1256082, and by a University of California President's Postdoctoral Fellowship. This work was facilitated through the use of advanced computational, storage, and networking infrastructure provided by the Hyak supercomputer system at the University of Washington. We would also like to acknowledge high-performance computing support from Yellowstone (ark:/85065/d7wd3xhc) provided by NCAR's Computational and Information Systems Laboratory, sponsored by the National Science Foundation. This work was performed as part of the NASA Astrobiology Institute's Virtual Planetary Laboratory Lead Team, supported by the National Aeronautics and Space Administration through the NASA Astrobiology Institute under solicitation NNH12ZDA002C and Cooperative Agreement Number NNA13AA93A. The authors wish to thank Dorian Abbot and an anonymous referee for their comments and suggestions, which improved the paper. B.C. acknowledges support from an appointment to the NASA Postdoctoral Program at NAI Virtual Planetary Laboratory, administered by Oak Ridge Affiliated Universities. R.B. acknowledges support from NSF grant  AST-1108882. A.S. thanks Brad Hansen, Jonathan Mitchell, John Johnson, Russell Deitrick, and Tom Quinn for helpful discussions regarding this work. 

\section{Author Disclosure Statement}

No competing financial interests exist.
\newpage
 \renewcommand{\theequation}{A\arabic{equation}}
  % redefine the command that creates the equation no.
  \setcounter{equation}{0}  % reset counter 
    \begin{center}
      {\bf APPENDIX\\ Determining Initial Planet Locations}
    \end{center}

	The location of a planet can be expressed in terms of the planet's true anomaly, the longitude measured from the direction of the pericenter of the planet's orbit. We used the transit times for each of the five Kepler-62 planets and calculated the true anomaly values for all five planets at the same epoch, using one planet's location as a reference point. 
	
	As stated in Section 2.1.1, we do not know the eccentricity of Kepler-62f (we assumed $e=0$ for Kepler-62b-e). The angular velocity and location of a planet in its orbit depends on its eccentricity, following Kepler's second law; therefore, we do not know the pericenter of the orbit for Kepler-62f. We therefore calculated the true anomaly for a range of possible eccentricities for Kepler-62f. Here we outline the equations used in our model to generate the true anomaly values, given assumed values for other key orbital parameters. 
\begin{center}
{\it Time of pericenter passage}
 \end{center}
	The time $t(f_i)$ that it takes for a planet to go from some initial reference point (to which it arrives at $t_o$) to the point of closest approach (where it arrives at $t_{peri}$), can be calculated using the following formula given by Sudarsky \emph{et al.} (\citeyear{Sudarsky2005}):
\begin{equation}\label{sud}
t(f_i)=\frac{-(1-e^2)^{1/2}P}{2\pi}\left \{\frac{e\sin f_i}{1+e\cos f_i}-2(1-e^2)^{-1/2} \tan^{-1}\left [\frac{(1-e^2)^{1/2}\tan(f_i/2)}{1+e}\right]\right\}.
\end{equation}
Taking the relation that $t(f_i)=t_o-t_{peri}$, we rearranged \eqref{sud} to yield an expression for $t_{peri}$:
\begin{equation}\label{sud2}
t_{peri}=t_o+\frac{(1-e^2)^{1/2}P}{2\pi}\frac{e\sin f_i}{1+e\cos f_i}-\frac{P}{\pi}\tan^{-1}\left[\frac{(1-e^2)^{1/2}\tan(f_i/2)}{1+e}\right].
\end{equation}

	To find each planet's true anomaly, it is necessary to first determine the location of each planet when it transits its star, $f_i$. Using Figure 1 for reference, with $\Theta=f+\omega=0$ when the planet $m_2$ passes through the sky plane (the plane perpendicular to the reference direction), at the time the planet passes behind the star, $\Theta=\pi/2$. By extension, the planet is in front of the star and transiting at $\Theta=3\pi/2$. Therefore, for any value of $\omega$, $f_i=3\pi/2-\omega$. Since we have assumed $\omega=0$ for all planets except Kepler-62f, $f_i=3\pi/2$ for these planets. Values of $f_i$ for Kepler-62f varied depending on the value of $\omega$. Values used in Equation \eqref{sud2} for the orbital period $P$, the transit time for each planet $t_o$, $e$, and $f_i$ are given in Table 6. With these values, $t_{peri}$ was calculated for each planet.

\begin{center}	
{\it From Mean Anomaly to True Anomaly}
\end{center}
The location of a planet in its orbit around a central star is generally described by one of three angular parameters: the \emph{mean anomaly}, the \emph{eccentric anomaly}, and the \emph{true anomaly}. The mean anomaly is a function of the average angular velocity of the planet, and therefore does not provide a precise location. Rather, the mean anomaly is an approximate location that is easy to calculate:
\begin{equation}\label{M}
M=n(t_o-t_{peri}),
\end{equation}
where $n=2\pi/P$. We used the first transit time for Kepler-62f as the reference point $t_o$ (see Table 6). 

The eccentric anomaly $E$ is an angle measured from a line through the focus of the planet's elliptical orbit to the radius of a circle that passes through the pericenter of the ellipse. It is related to the mean anomaly and the orbital eccentricity $e$ through Kepler's equation:
\begin {equation}\label{kepler}
M=E-esinE.
\end{equation}

As this equation cannot be solved analytically, we solved it numerically following the method provided by Danby and Burkardt (\citeyear{Danby1983}). With calculated values of $E$ for our assigned eccentricities, the true anomaly was calculated using the following relation:
\begin{equation}\label{true}
f=2tan^{-1}\left[\left(\frac{1+e}{1-e}\right)^{1/2}tan(E/2)\right].
\end{equation}

Eq. \eqref{true} allows us to calculate the relative positions of all the planets at any given time so that our \emph{n}-body integrations will realistically reproduce the system's orbital evolution.

\newpage
%\bibliography{/Users/aomawashields/Dropbox/AAPF/research/refs.bib}

\newpage
%%%%%%%%%%%%%%%%%%%%%%%%%%%%%%
%   Tables
%%%%%%%%%%%%%%%%%%%%%%%%%%%%%%
\linespread{1.0}
\begin{sidewaystable}[!htp] 
\caption{Orbital elements used to describe Kepler-62 planetary orbits. Angles here are in radians except for the orbital inclination, which is listed in degrees. In exoplanet geometry, $i=90^\circ$ constitutes an edge-on orbit capable of yielding a transit observable by the Earth. Data for the semi-major axis and the orbital inclination are from Borucki \emph{et al.} (\citeyear{Borucki2013}), though we have omitted the uncertainties here for ease of reading.} 
\vspace{2 mm}
\centering \begin{tabular}{c|l|c} 
\hline\hline 
 Parameters & Definition & Initial Values (Kepler-62b, 62c, 62d, 62e, 62f)\\  [0.5ex]
%heading 
\hline
Semi-major Axis $a$ (AU) & Half of the longest diameter of an orbit & 0.0553, 0.0929, 0.120, 0.427, 0.718\\ \hline
Eccentricity $e$ & The degree of ellipticity (oval shape) of an orbit \\ & ($e=0$ - circular; $e=1$ - parabolic) & 0.0, 0.0, 0.0, 0.0, 0.0\textendash 0.9\\ \hline
Inclination $i$ ($^\circ$)& The angle between the planet's orbital plane \\ & and the sky plane & 89.2, 89.7, 89.7, 89.98, 89.90\\ \hline
Longitude of Ascending Node $\Omega$ & The angle between the sky plane's 0$^\circ$ longitude\\ & and the point where the planet passes  from in \\ & front of to behind the star &  0.0, 0.0, 0.0, 0.0, 0.0\\ \hline
Longitude of Pericenter $\omega$ & The angle between a reference direction \\ & and the point of closest approach of a \\ & planet to its star (pericenter) & 0.0, 0.0, 0.0, 0.0, 0.0 $\rightarrow2\pi$ ($\Delta\omega=\pi/6$)\\ \hline
True Anomaly $f$ & The angle between a radius vector through \\ & pericenter and the planet's location \\ & on an orbital ellipse around its host star & Calculated in Appendix\\  [1ex]
\hline 
\end{tabular} 
\label{table:nonlin} 
\end{sidewaystable}
\pagebreak

\linespread{1.0}
\begin{table}[!htp] 
\caption{Masses (M$_\oplus$) used as inputs to HNBody for different sets of orbital integrations of the planets Kepler-62b-f.} 
\vspace{2 mm}
\centering \begin{tabular}{c c c c c c} 
\hline\hline 
Kepler-62b & Kepler-62c & Kepler-62d & Kepler-62e & Kepler-62f & Relation/Source \\  [0.5ex]
%heading 
\hline
2.3 & 0.1 & 8.2 & 4.4 & 2.9 &  Kopparapu  \emph{et al.} (\citeyear{Kopparapu2014})\\ 
2.8 & 0.1 & 5.0 & 4.2 & 3.7 &  Weiss  \emph{et al.} (\citeyear{Weiss2014})\\
9.0 & 4.0 & 14 & 36 & 35 &  Borucki \emph{et al.} (\citeyear{Borucki2013})\\[1ex]
\hline 
\end{tabular} 
\label{table:nonlin} 
\end{table}
\clearpage
\pagebreak

\linespread{1.0}
\begin{table}[!htp] 
\caption{CAM4 spectral wavelength bands specifying shortwave (stellar) incoming flux into the atmosphere, and the percentage of flux within each waveband for a synthetic spectrum of a K-dwarf star with a similar photospheric temperature to Kepler-62, from the Pickles Stellar Atlas \citep{Pickles1998}.} 
\vspace{2 mm}
\centering \begin{tabular}{c c c c c c} 
\hline\hline 
Band & $\lambda_{min}$ & $\lambda_{max}$ & K2V star  \% flux\\  [0.5ex]
%heading 
\hline
$1$ & 0.200 & 0.245 & 0.128\\
$2$ & 0.245 & 0.265 & 0.075\\
$3$ & 0.265 & 0.275 & 0.054\\
$4$ & 0.275 & 0.285 & 0.056\\
$5$ & 0.285 & 0.295 & 0.070\\
$6$ & 0.295 & 0.305 & 0.091\\
$7$ & 0.305 & 0.350 & 1.076\\
$8$ & 0.350 & 0.640 & 27.33\\
$9$ & 0.640 & 0.700 & 6.831\\
$10$ & 0.700 & 5.000 & 64.35\\ 
$11$ & 2.630 & 2.860 & 0.000\\ 
$12$ & 4.160 & 4.550 & 0.000\\ [1ex]
\hline 
\end{tabular} 
\label{table:nonlin} 
\end{table}
\pagebreak

\linespread{1.0}
\begin{sidewaystable}[!htp] 
\caption{Additional input parameters used in CCSM4 climate simulations of Kepler-62f. Here the parameter ``VEP" is the angle of the vernal equinox relative to pericenter. Run7 is the synchronous rotation rate case, where the planet's rotation period is equal to its orbital period.} 
\vspace{2 mm}
\centering \begin{tabular}{c c c c c c c c} 
\hline\hline 
Parameter & Run1 & Run2 & Run3 & Run4 & Run5 & Run6 & Run7\\  [0.5ex]
\hline
Orbital eccentricity & 0.0 & 0.1 & 0.2 & 0.32 & 0.32 &0.32 & 0.0\\
Obliquity ($^\circ$) & 23 & 23 & 23 & 23 & 60 & 23 & 0\\
Rotation rate & 24 hrs & 24 hrs & 24 hrs & 24 hrs & 24 hrs & 24 hrs & 365 days\\
VEP ($^\circ$) & 90 & 90 & 90 & 90 & 90 & 0 & 90\\[1ex]
\hline 
\end{tabular} 
\label{table:nonlin}
\end{sidewaystable}
\pagebreak

\linespread{1.0}
\begin{sidewaystable}[!htp] 
\caption{Additional input parameters used in high-CO$_2$ LMD Generic GCM climate simulations of Kepler-62f. Two additional simulations were run with Earth-like CO$_2$, obliquities of 23$^\circ$ and 60$^\circ$, $e=0.32$ and a VEP of 90$^\circ$, for comparison with the CCSM4 simulations. Both sets of simulations are shown in Figure 12.} 
\vspace{2 mm}
\centering \begin{tabular}{c c c c c c c c c c c c c} 
\hline\hline 
Parameter & Run1 & Run2 & Run3 & Run4 & Run5 & Run6 & Run7 & Run8 & Run9 & Run10 & Run11 & Run12-27\\[0.5ex]
%heading 
\hline
Ecc & 0.32 & 0.00 & 0.32 & 0.32 & 0.32 & 0.32 & 0.32 & 0.00 & 0.00 & 0.00 & 0.00 & 0-0.32\\
Obl ($^\circ$) & 23.5 & 23.5 & 23.5 & 23.5 & 23.5 & 23.5 & 23.5 & 0.00 & 0.00 & 0.00 & 0.00 & 0-90\\
CO$_2$ (bars)       & 1 & 3 & 3 & 5 & 8 & 10 & 12 & 1 & 1 & 3 & 3 & 3\\
Rot & 24 hrs & 24 hrs & 24 hrs & 24 hrs & 24 hrs & 24 hrs & 24 hrs & 24 hrs & 267 days & 24 hrs & 267 days & 24 hrs\\
VEP & 0 & 0 & 0 & 0 & 0 & 0 & 0 & 0 & 0 & 0 & 0 & 0-90\\[1ex]
\hline 
\end{tabular} 
\label{table:nonlin} 
\end{sidewaystable}
\pagebreak

\linespread{1.0}
\begin{table}[!htp] 
\caption{Key parameters used as inputs to Eq. \eqref{sud2} for Kepler-62b-f. BJD (Barycentric Julian Date) is the Julian date (the number of days since the beginning of the Julian period, ca. 4713 BC) corrected for Earth's changing position relative to the center of mass (barycenter) of the Solar System.} 
\vspace{2 mm}
\centering \begin{tabular}{c c c c c c} 
\hline\hline 
 & Kepler-62b & Kepler-62c & Kepler-62d & Kepler-62e & Kepler-62f\\  [0.5ex]
%heading 
\hline
$P$ (days) & 5.714932 & 12.4417 & 18.16406 & 122.3874 & 267.291 \\ 
$t_o$ (BJD-2454900) & 103.9189 & 67.651 & 113.8117 & 83.404 & 522.710 \\
$e$ & 0.0 & 0.0 & 0.0 & 0.0 & 0.0\textendash 0.9\\ 
$f_i$ & $3\pi/2$ & $3\pi/2$ & $3\pi/2$ & $3\pi/2$ & $-\pi/2$ $\rightarrow$ $3\pi/2$\\ [1ex]
\hline 
\end{tabular} 
\label{table:nonlin} 
\end{table}
\clearpage
\pagebreak
%%%%%%%%%%%%%%%%%%%%%%%%%%%%%%
%    Figures
%%%%%%%%%%%%%%%%%%%%%%%%%%%%%%
\linespread{1.15}
\begin{figure}[!htb]
\begin{center}
\includegraphics [scale=0.75]{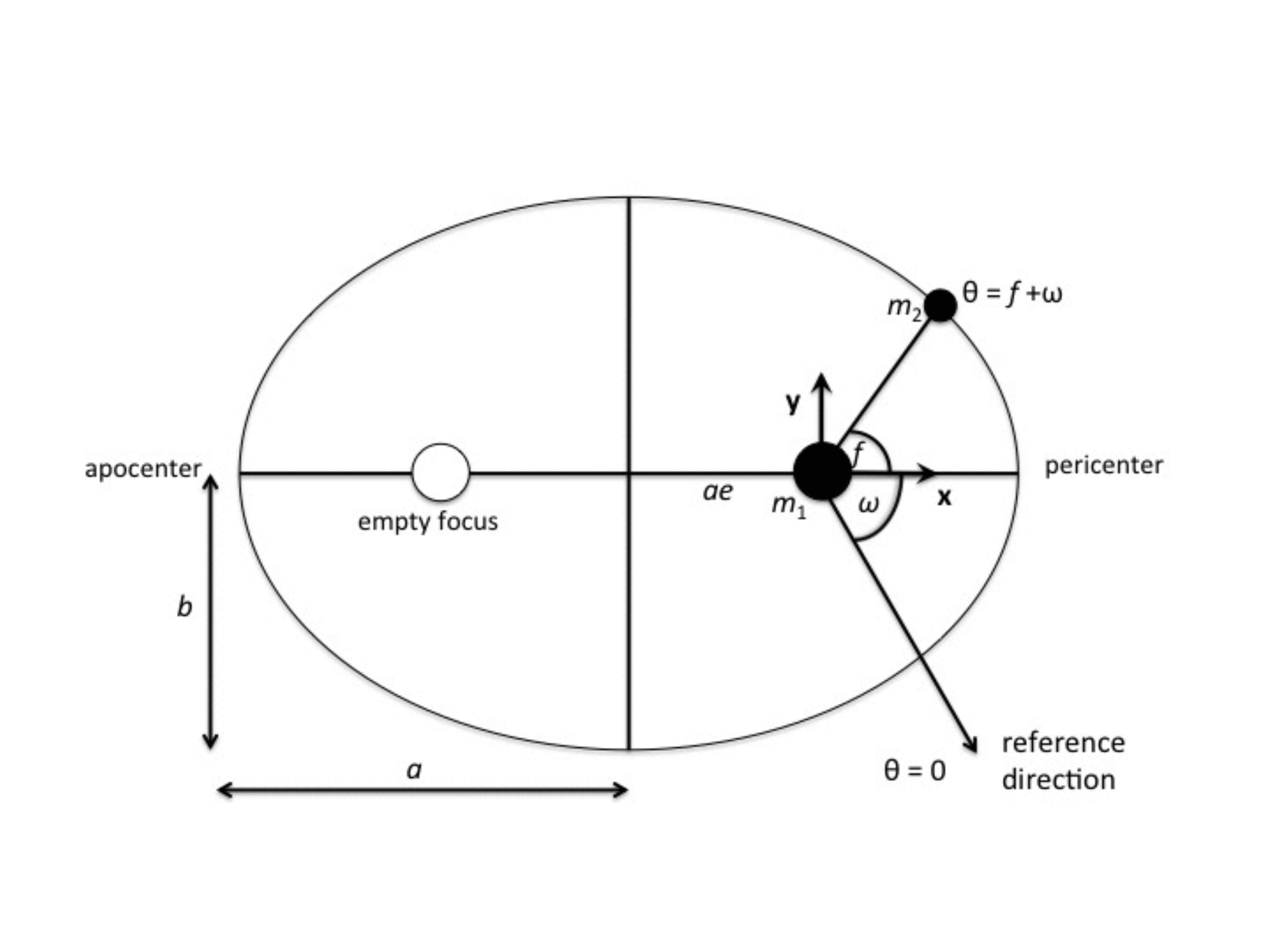}\\
\caption{Geometry of the elliptical orbit of a body of mass $m_2$ around $m_1$. The ellipse has semi-major axis $a$, semi-minor axis $b$, eccentricity $e$, and longitude of pericenter $w$. The true anomaly $f$ denotes the angle subtended by an imaginary line connecting $m_1$ with the location of $m_2$ in its orbit and one connecting $m_1$ with pericenter (the point of closest approach to $m_1$). This assumes that $m_1$ $\gg$  $m_2$.}
\label{Figure 1. }
\end{center}
\end{figure}
\pagebreak

\begin{figure}[!htb]
\begin{center}
\includegraphics [scale=0.80]{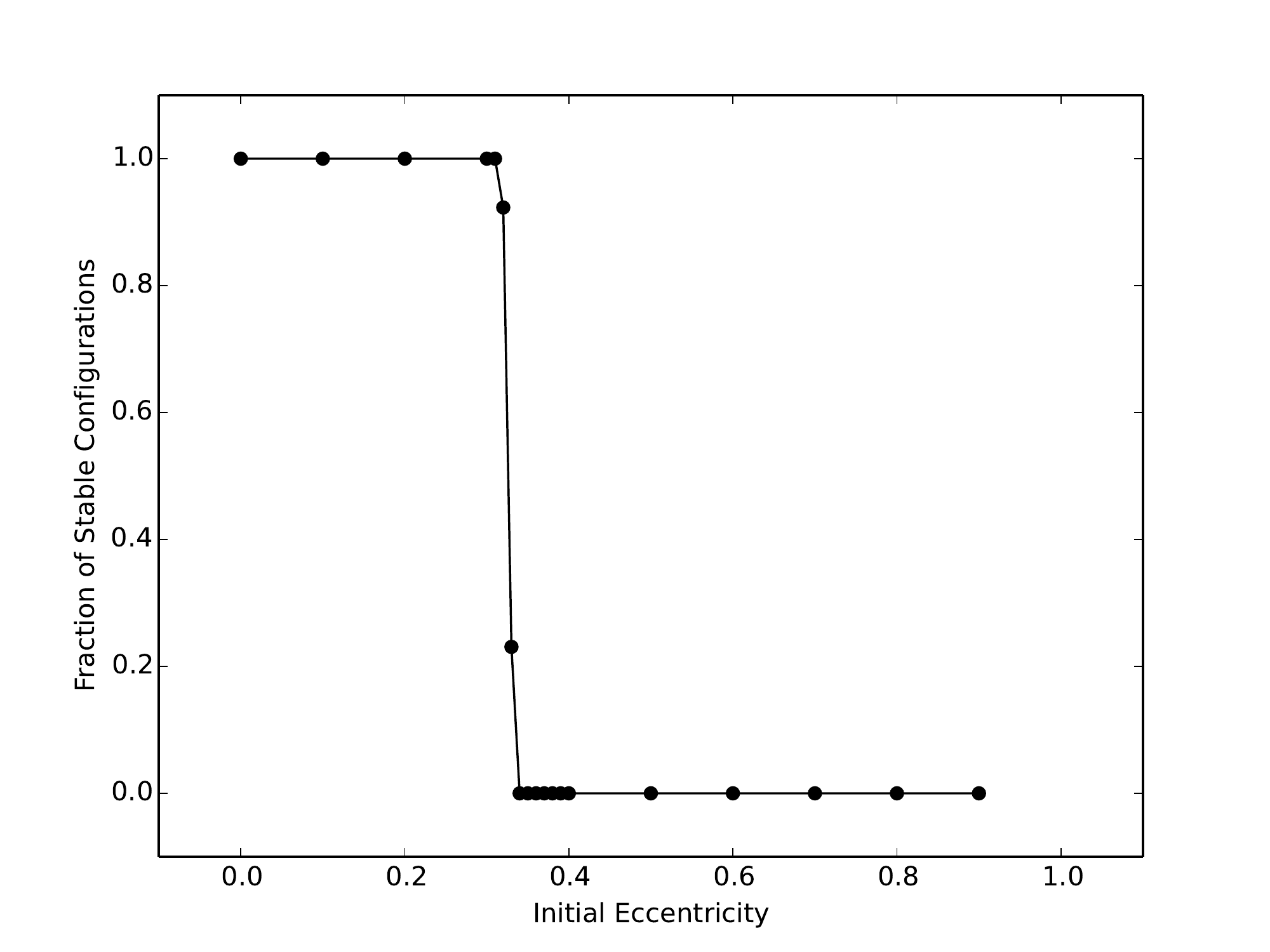}\\
\caption{Fraction of stable configurations after a $10^6$-yr HNBody integration for initial eccentricities between 0.0 and 0.9 for Kepler-62f, assuming the Kopparapu \emph{et al.} (\citeyear{Kopparapu2014}) mass-radius relationship. The eccentricities of all other planets in the Kepler-62 system were set to zero.}
\label{Figure 2.}
\end{center}
\end{figure}
\pagebreak

\begin{figure}[!htb]
\begin{center}
\includegraphics [scale=1.20]{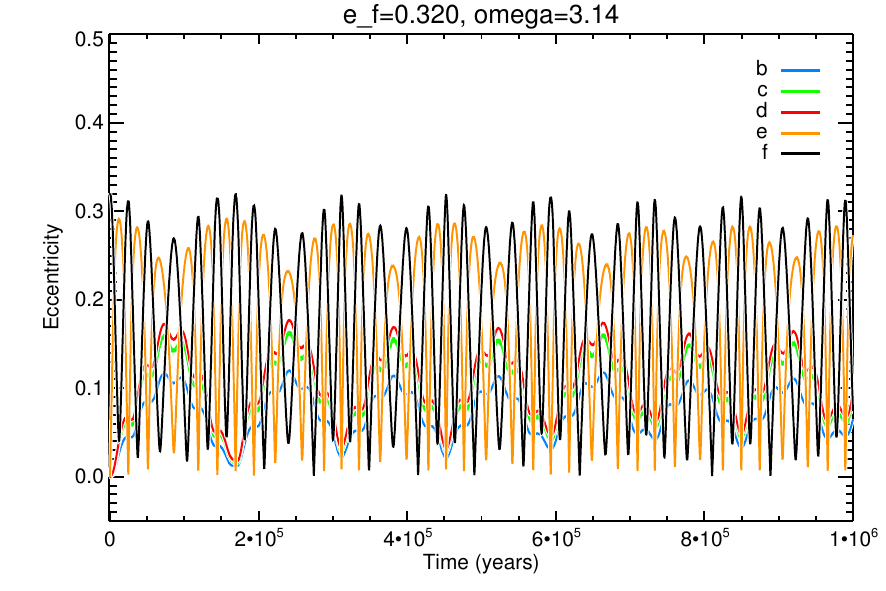}\\
\caption{Evolution of the eccentricities of Kepler-62b-f as a function of time calculated with HNBody. Initial eccentricities and longitudes of pericenter for Kepler-62b-e were set to zero. The initial eccentricity of Kepler-62f was set to 0.32 with a longitude of pericenter equal to $\pi$.}
\label{Figure 3.}
\end{center}
\end{figure}
\pagebreak

\begin{figure} 
\begin{tabular}{cc}
\includegraphics[width=0.49\textwidth]{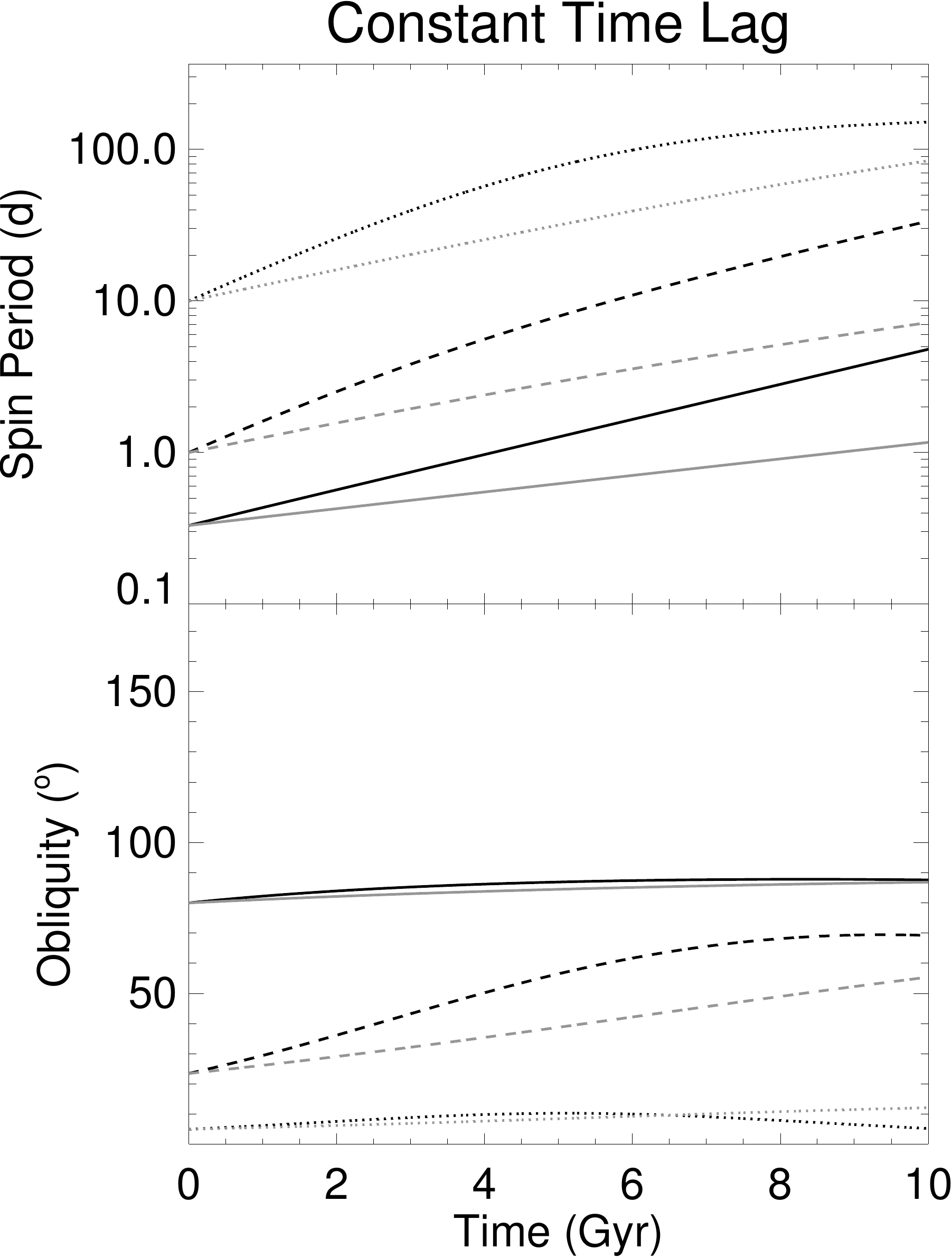} &
\includegraphics[width=0.49\textwidth]{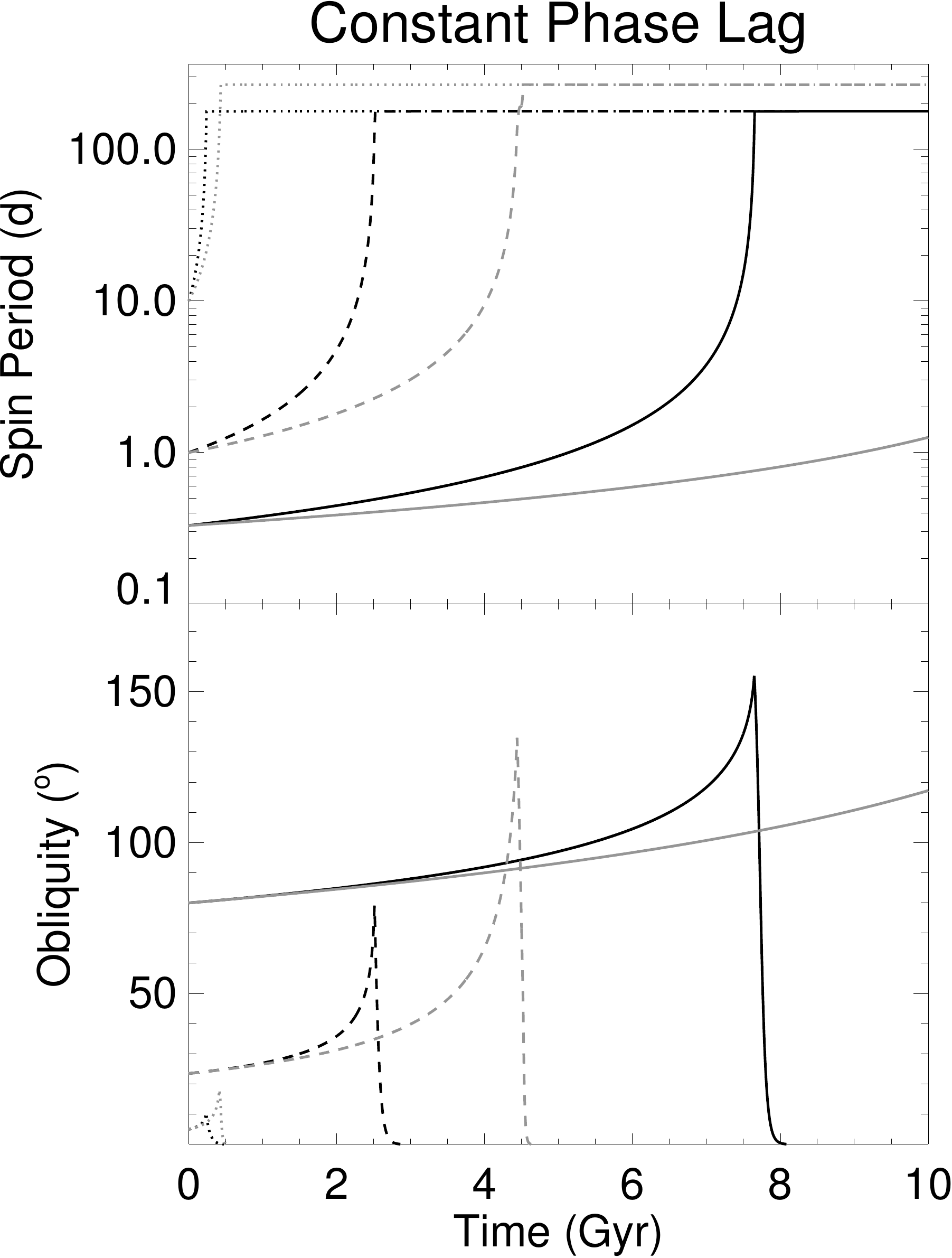}
\end{tabular}
\caption{Rotational evolution of Kepler-62f due to tidal processes. {\it Left:} Evolution of the spin period (top) and obliquity (bottom) for the CTL model. The gray curves assume a circular orbit, and the black curves assume an eccentricity of 0.32. Solid lines represent planets that begin with an 8-hr rotation period and an obliquity of $80^\circ$. Dashed lines assume the planet began with the modern Earth's rotational state. Dotted lines assume an initial spin period of 10 days and an obliquity of $5^\circ$. {\it Right:} Same as left panels, but for the CPL model.} 
\label{Figure 4}
\end{figure}
\pagebreak

\begin{figure}[!htb]
\begin{center}
\includegraphics [scale=0.70]{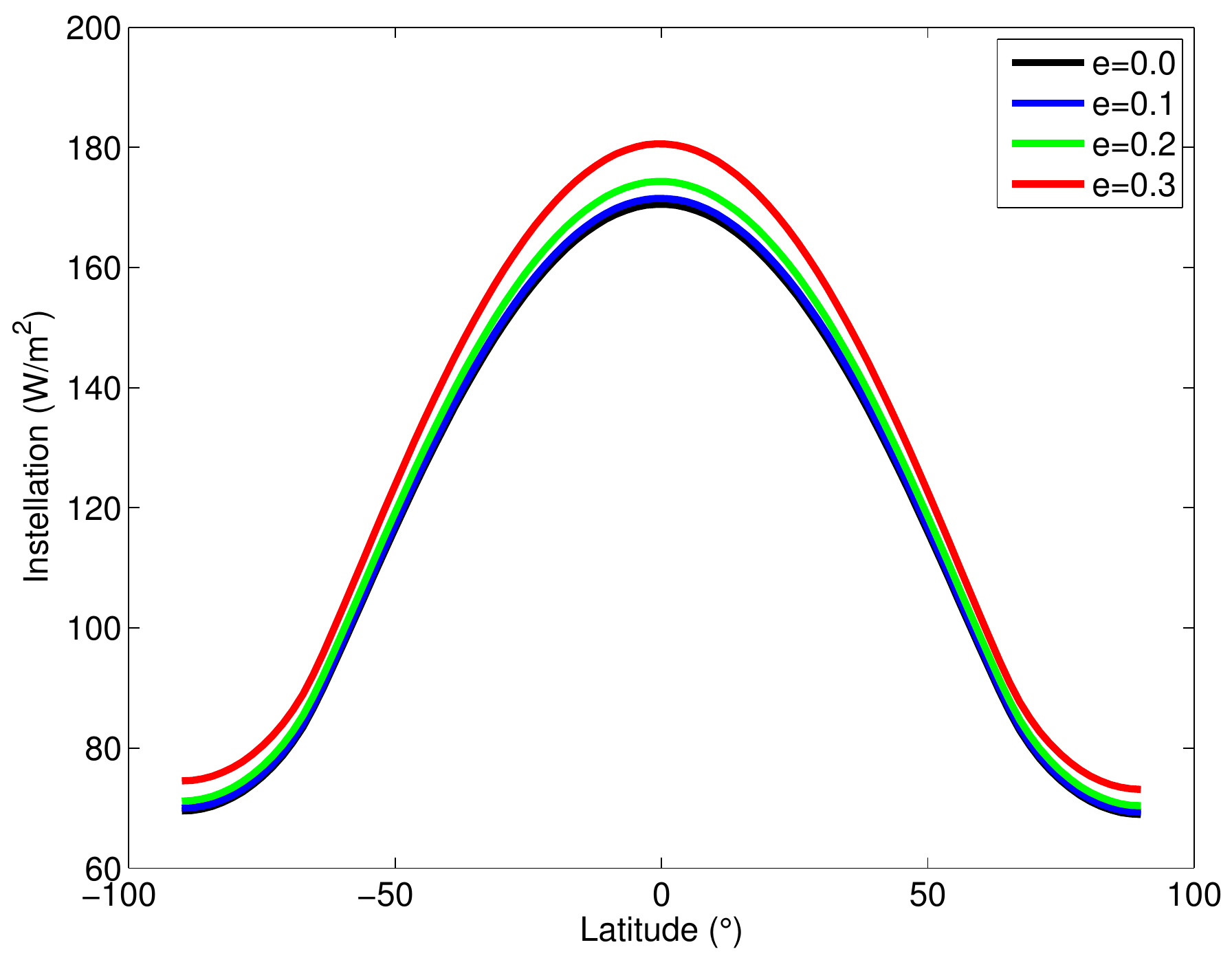}\\
\caption{Instellation as a function of latitude for Kepler-62f (using CCSM4), assuming $e=0.0$ (black), $e=0.1$ (blue), $e=0.2$ (green), and $e=0.32$ (red). The plots show 12-month averages. The obliquity of the planet was set to 23$^\circ$. The angle of the vernal equinox relative to pericenter was set to 90$^\circ$, similar to the Earth (102.7$^\circ$). The larger the eccentricity, the larger the annually-averaged instellation received at a given latitude.}
\label{Figure 5.}
\end{center}
\end{figure}
\pagebreak

\begin{figure}[!htb]
\begin{center}
\includegraphics [scale=2.00]{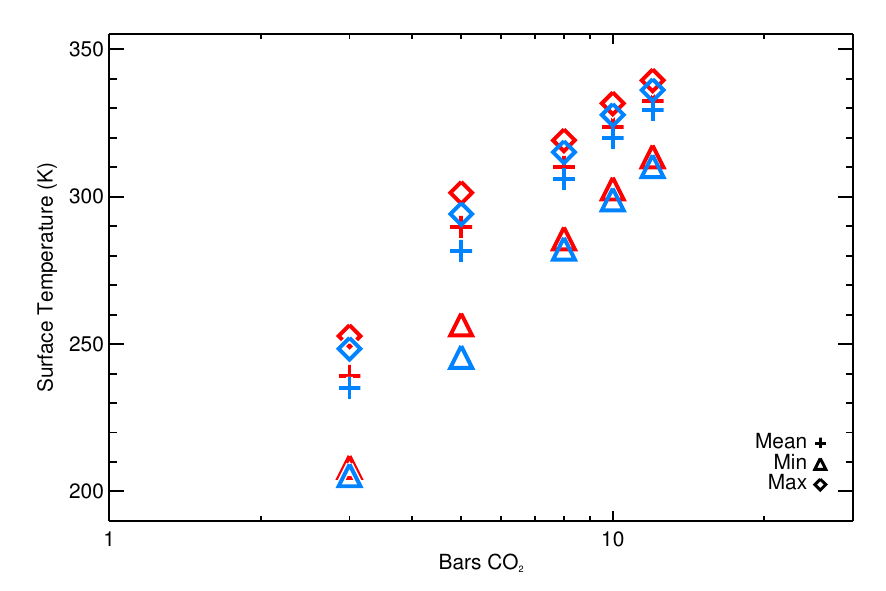}\\
\caption{Mean (plus symbols), minimum (triangles), and maximum (diamonds) surface temperature for Kepler-62f after 100-200-year LMD Generic GCM simulations, assuming $e=0.00$ (blue) and $e=0.32$ (red), an Earth-like rotation rate, and different levels of atmospheric CO$_2$. The mean values take into account the location of measured surface temperature values relative to the total surface area of the planet. An obliquity of 23.5$^\circ$ and VEP = 0$^\circ$ is assumed.}
\label{Figure 6.}
\end{center}
\end{figure}
\pagebreak

\begin{figure}[!htb]
\begin{center}
\includegraphics [scale=2.00]{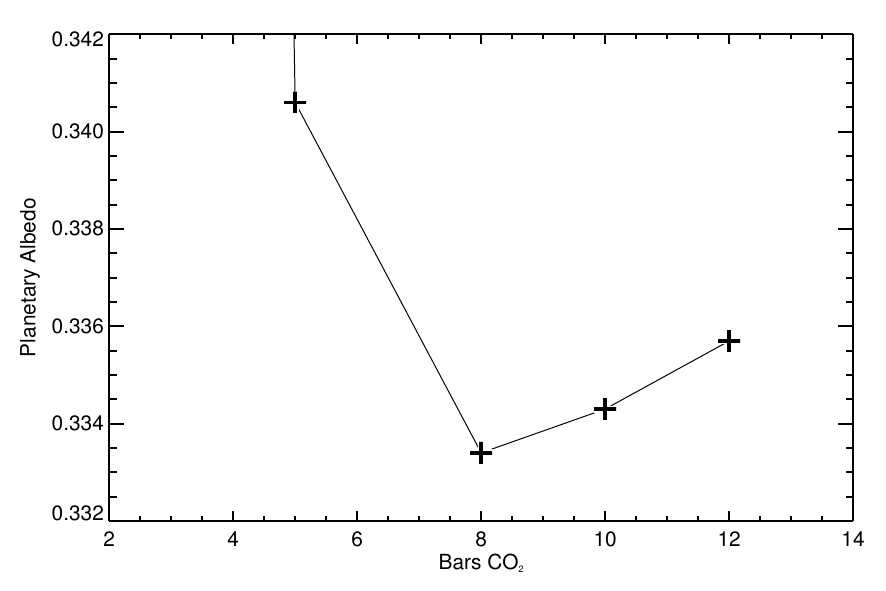}\\
\caption{Planetary albedo of Kepler-62f as a function of simulated atmospheric CO$_2$ concentration after 100-200-year LMD Generic GCM simulations. The 3-bar CO$_2$ simulation, which was ice-covered, with a planetary albedo of 0.491, is not shown, to enlarge the turning point in planetary albedo at 8 bars CO$_2$. An obliquity of 23.5$^\circ$, an eccentricity of zero, and VEP = 0$^\circ$ is assumed.}
\label{Figure 7.}
\end{center}
\end{figure}
\pagebreak

\begin{figure}[!htb]
\begin{center}
\includegraphics [scale=0.7]{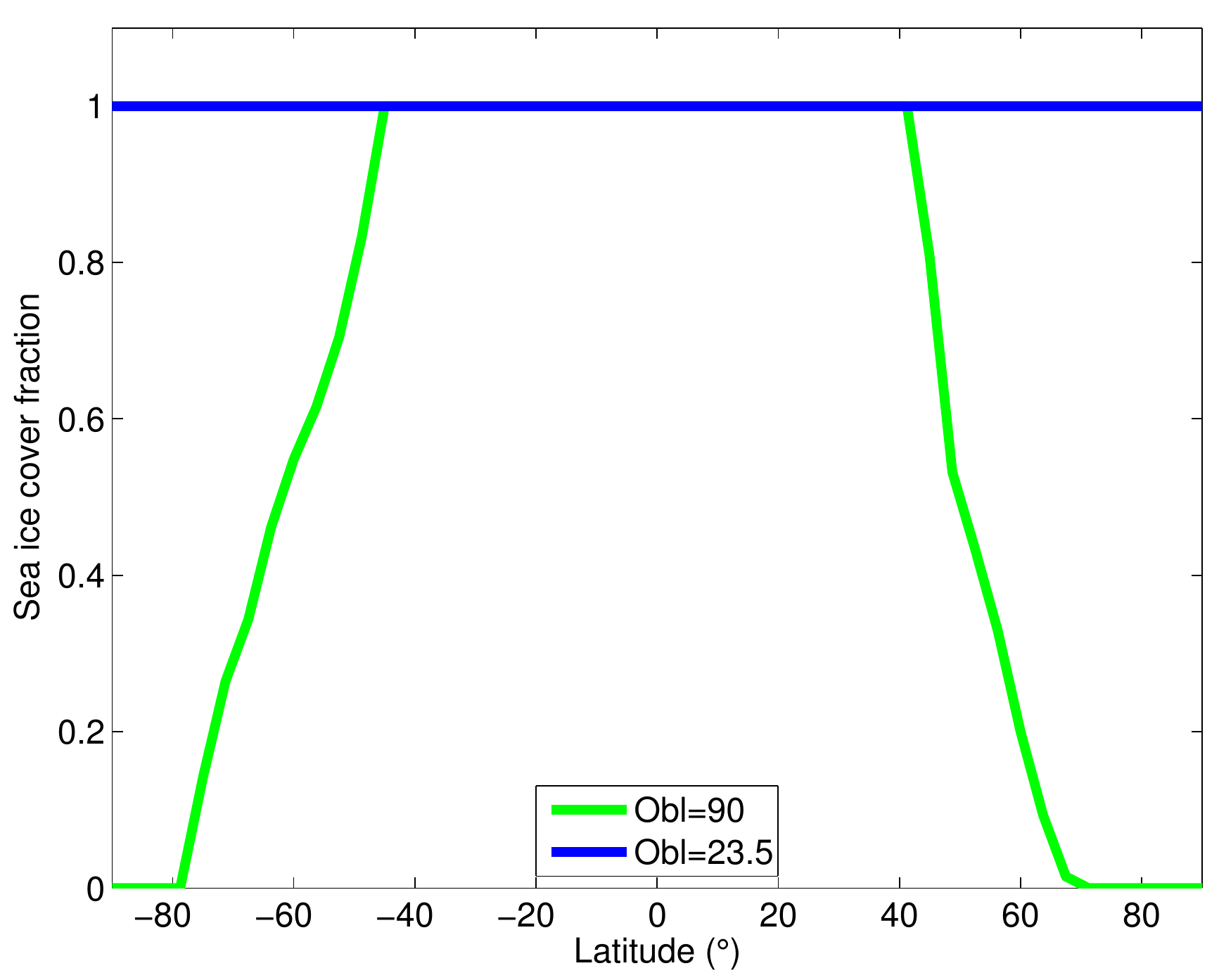}\\
\caption{Sea ice cover fraction as a function of latitude for Kepler-62f after 160-year LMD Generic GCM simulations, with 3 bars of CO$_2$ in the atmosphere, the maximum stable initial eccentricity ($e=0.32$), Earth-like (blue) and 90$^\circ$ (green) obliquities, and an Earth-like rotation rate. A VEP of 0$^\circ$ is assumed.}
\label{Figure 8.}
\end{center}
\end{figure}
\pagebreak

\begin{figure}[!htb]
\begin{center}
\includegraphics [scale=0.6]{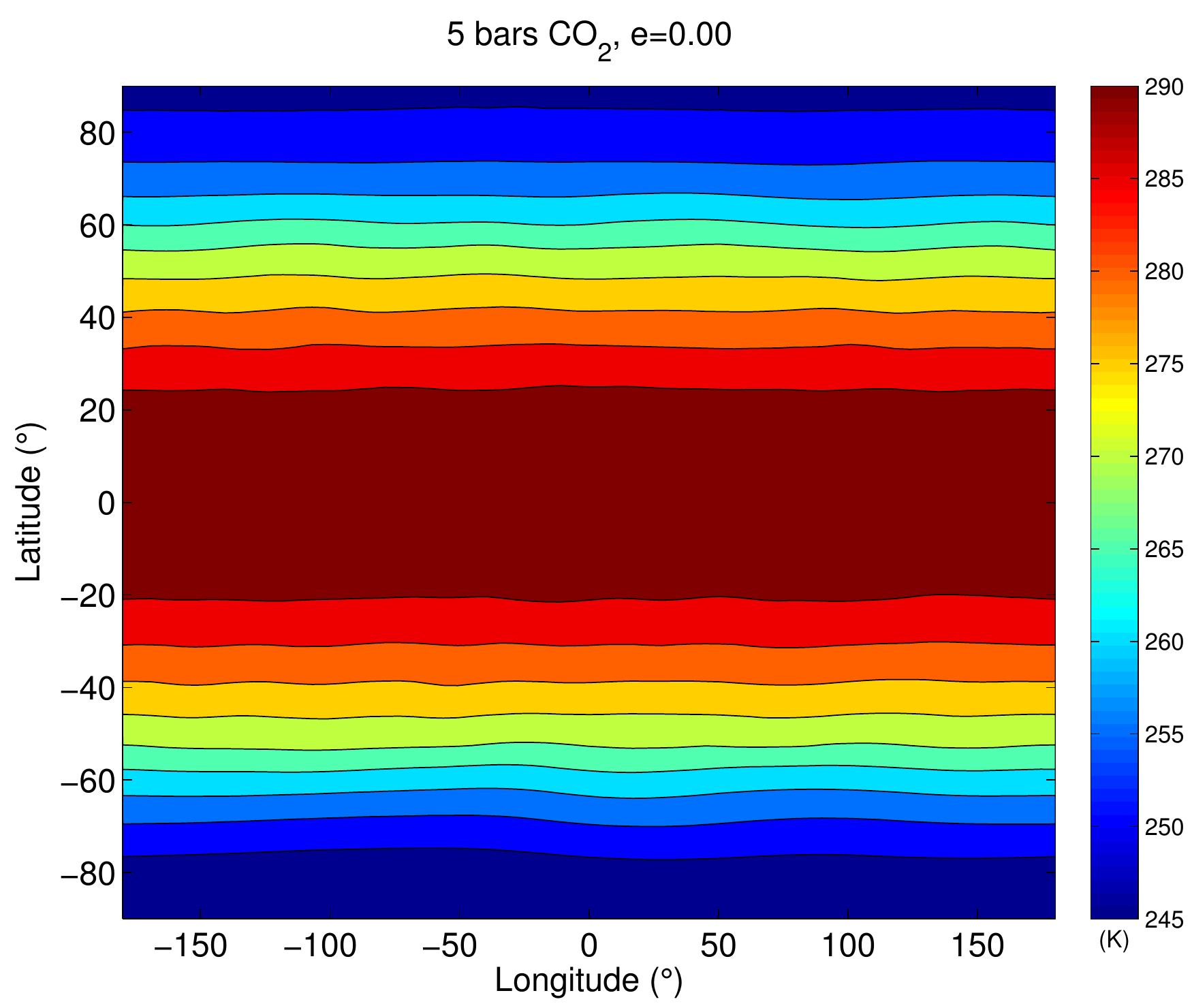}\\
\caption{Surface temperature as a function of latitude for Kepler-62f after a 120-year LMD Generic GCM simulation, assuming $e=0$, an obliquity of 23.5$^\circ$, 5 bars of CO$_2$, and an Earth-like rotation rate. A VEP of 0$^\circ$ is assumed.}
\label{Figure 9.}
\end{center}
\end{figure}
\pagebreak

\begin{figure}[!htb]
\begin{center}
\includegraphics [scale=0.4]{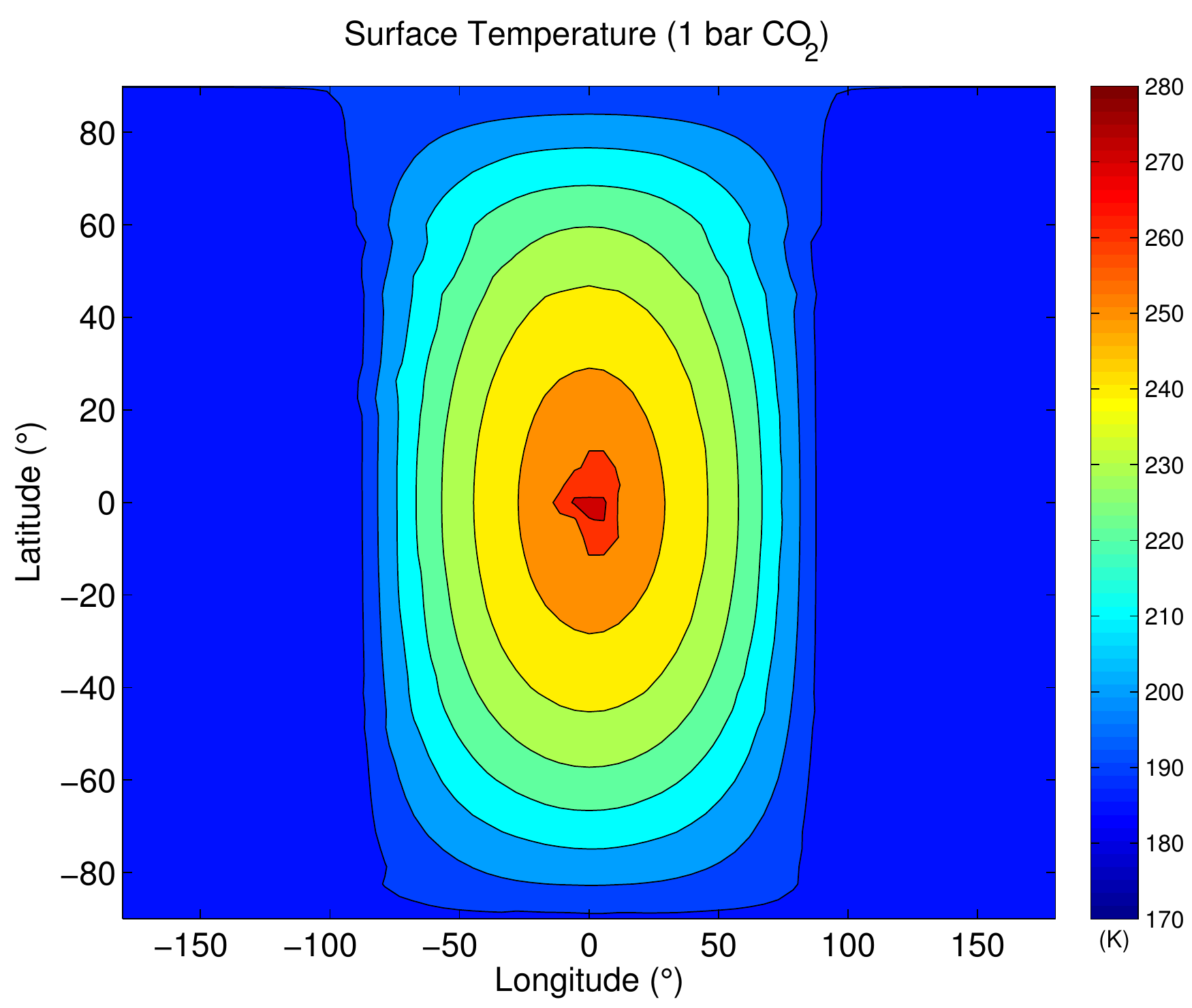}
\includegraphics [scale=0.4]{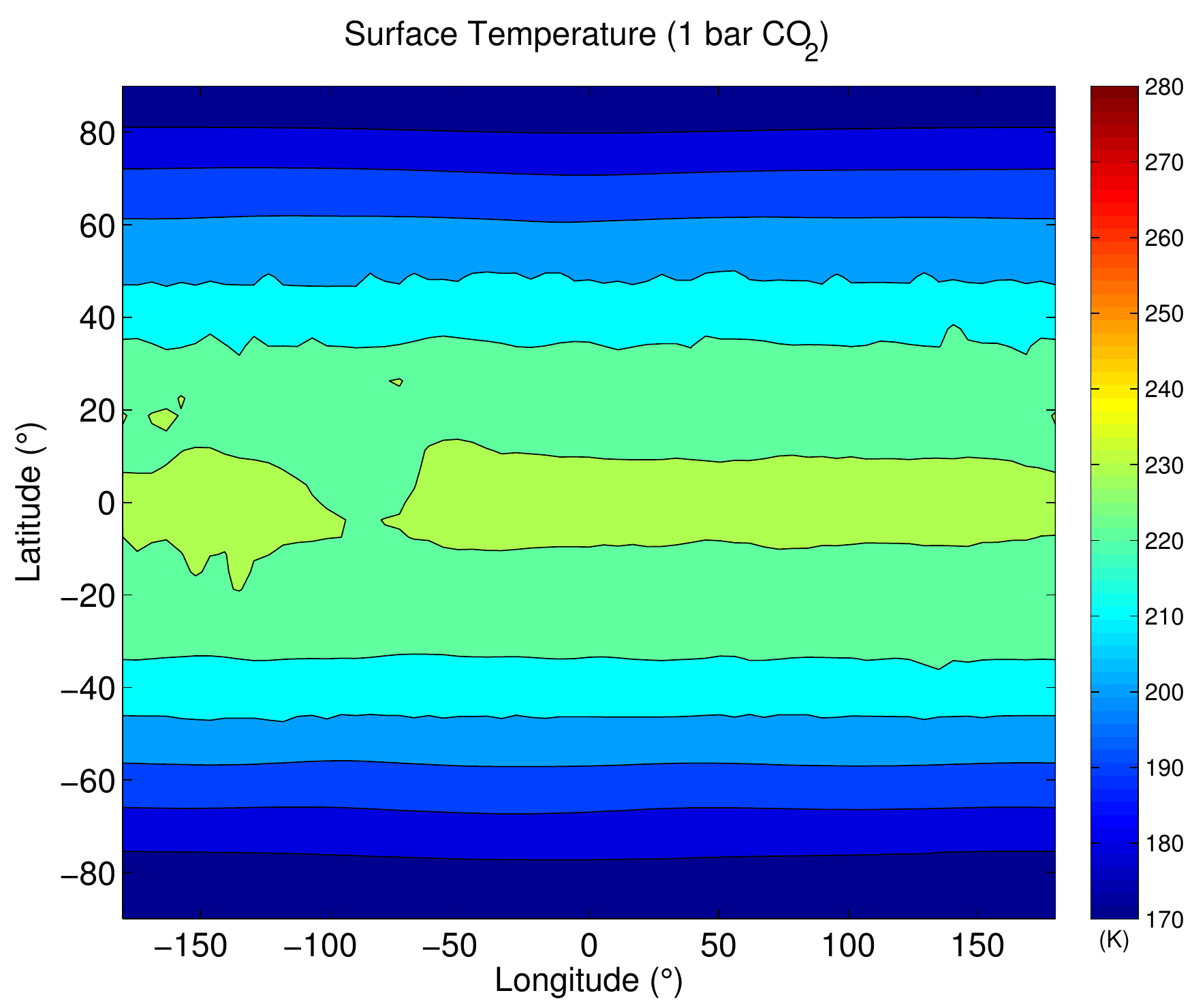}\\
\includegraphics [scale=0.4]{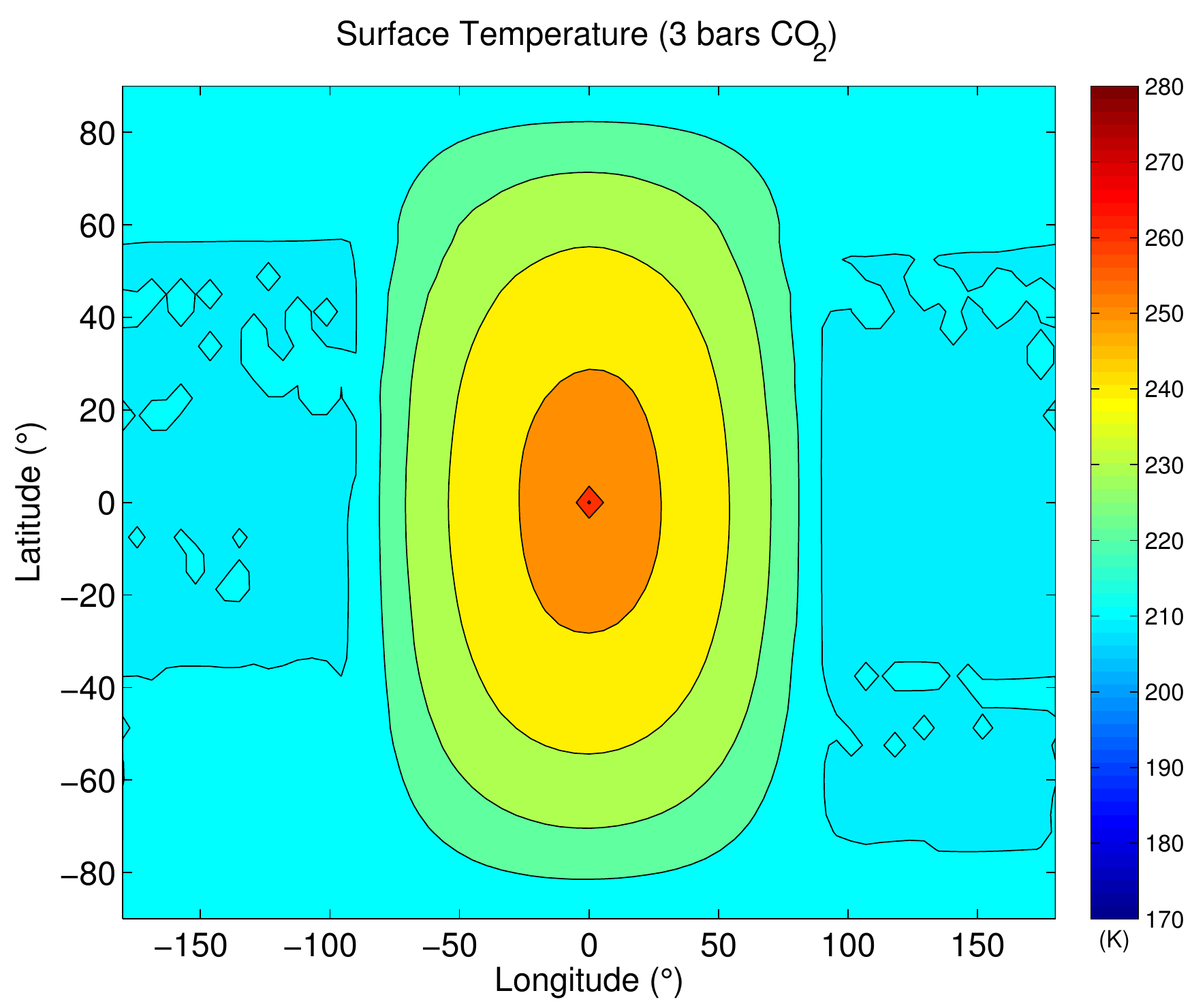}
\includegraphics [scale=0.4]{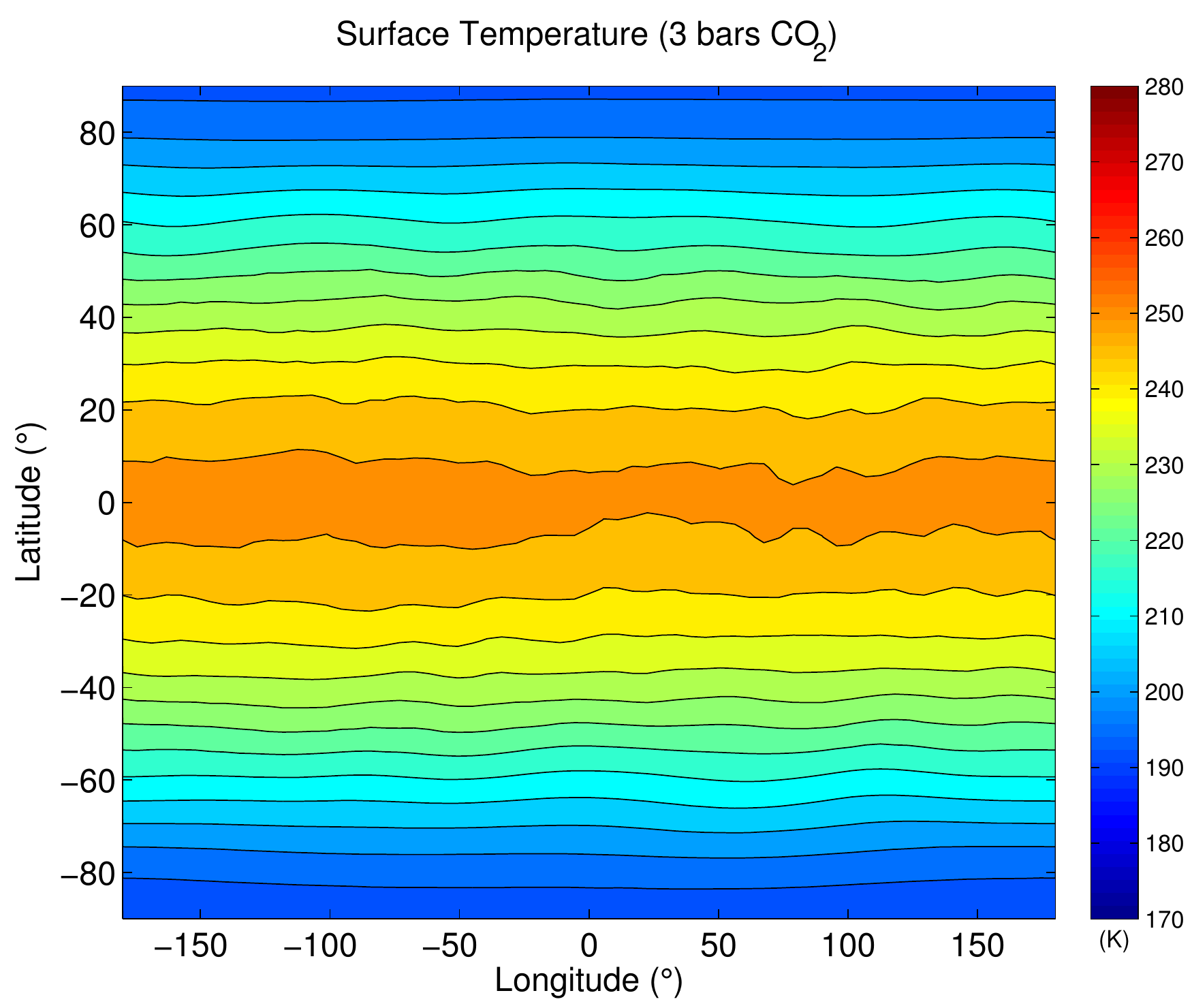}\\
\caption{Surface temperature for a synchronous (left) and Earth-like (24-hr) rotation rate for Kepler-62f, after a 65-100-year LMD Generic GCM simulation with 1 bar (top) and 3 bars (bottom) of CO$_2$ in the atmosphere. We assumed $e=0$, VEP = 0$^\circ$ and an obliquity of 0$^\circ$ for all four simulations.}
\label{Figure 10.}
\end{center}
\end{figure}
\pagebreak

\begin{figure}[!htb]
\begin{center}
\includegraphics [scale=0.4]{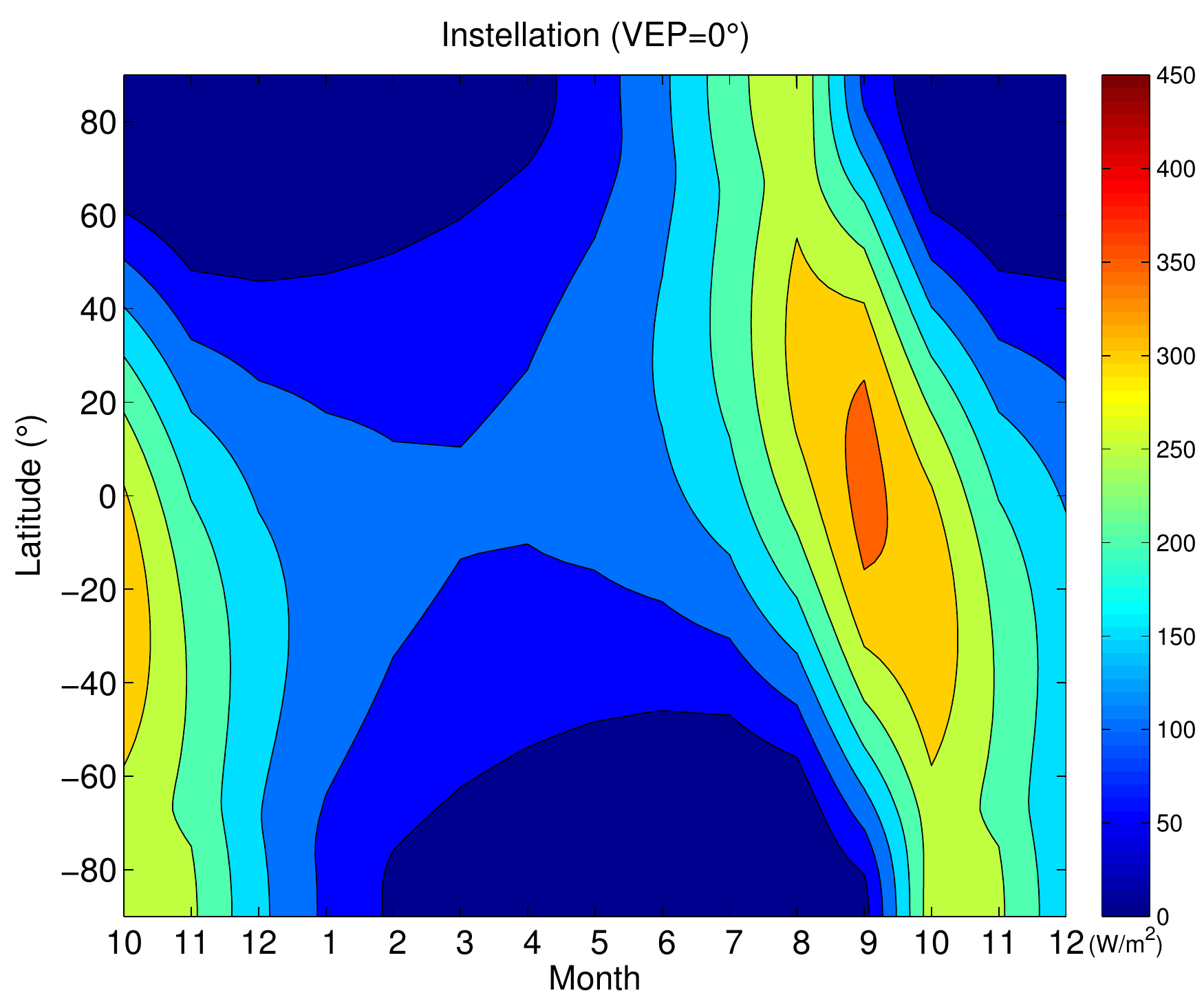}
\includegraphics [scale=0.4]{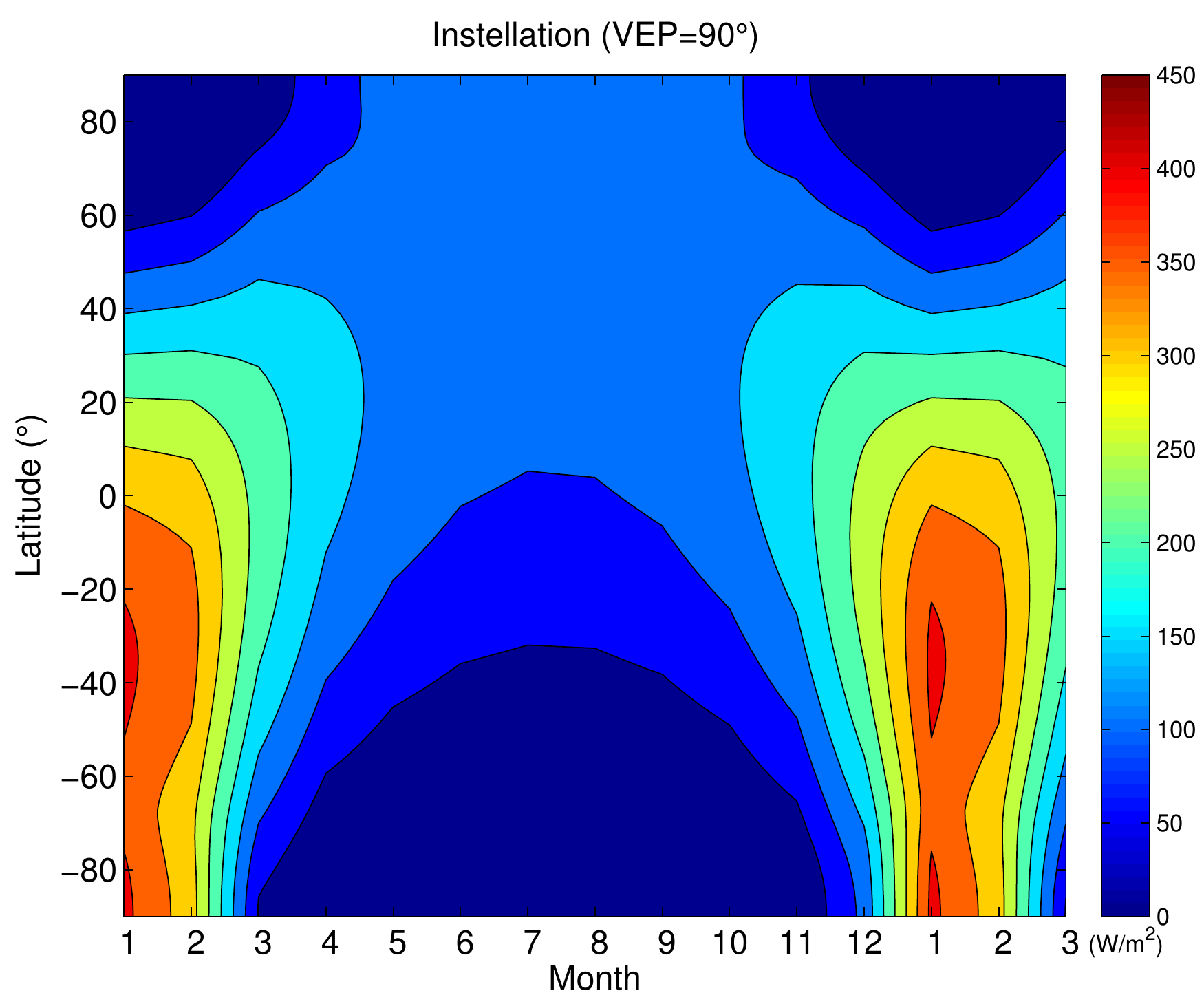}\\
\caption{Annual mean instellation as a function of latitude for Kepler-62f as a function of the month of the year after 40-yr CCSM4 simulations, assuming a 12-month annual cycle and a VEP of 0$^\circ$ (left), and 90$^\circ$ (right). The obliquity and eccentricity of the planet was set to 23$^\circ$ and 0.32, respectively.}
\label{Figure 11.}
\end{center}
\end{figure}
\pagebreak

\begin{figure}[!htb]
\begin{center}
\includegraphics [scale=0.4]{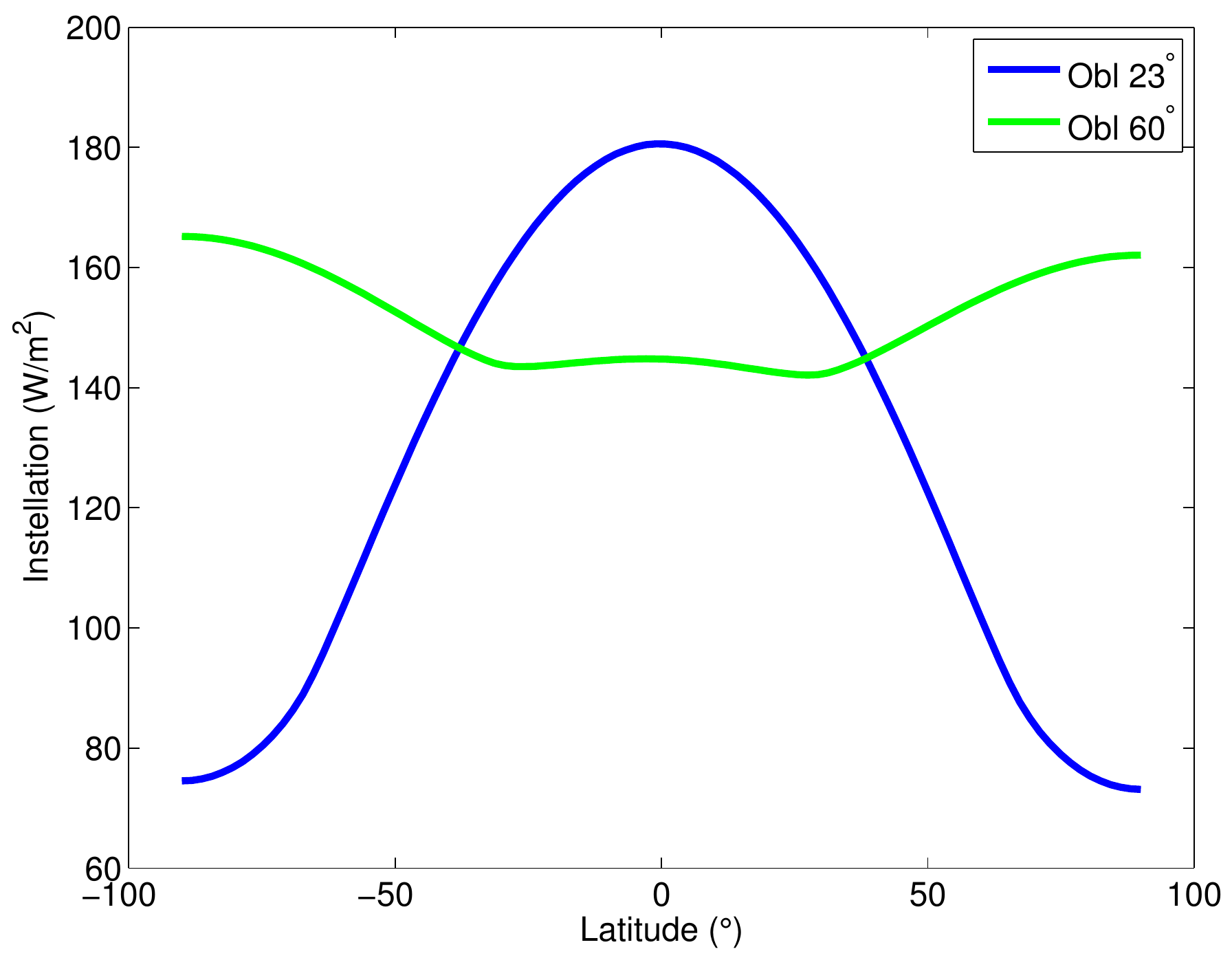}\\
\includegraphics [scale=0.4]{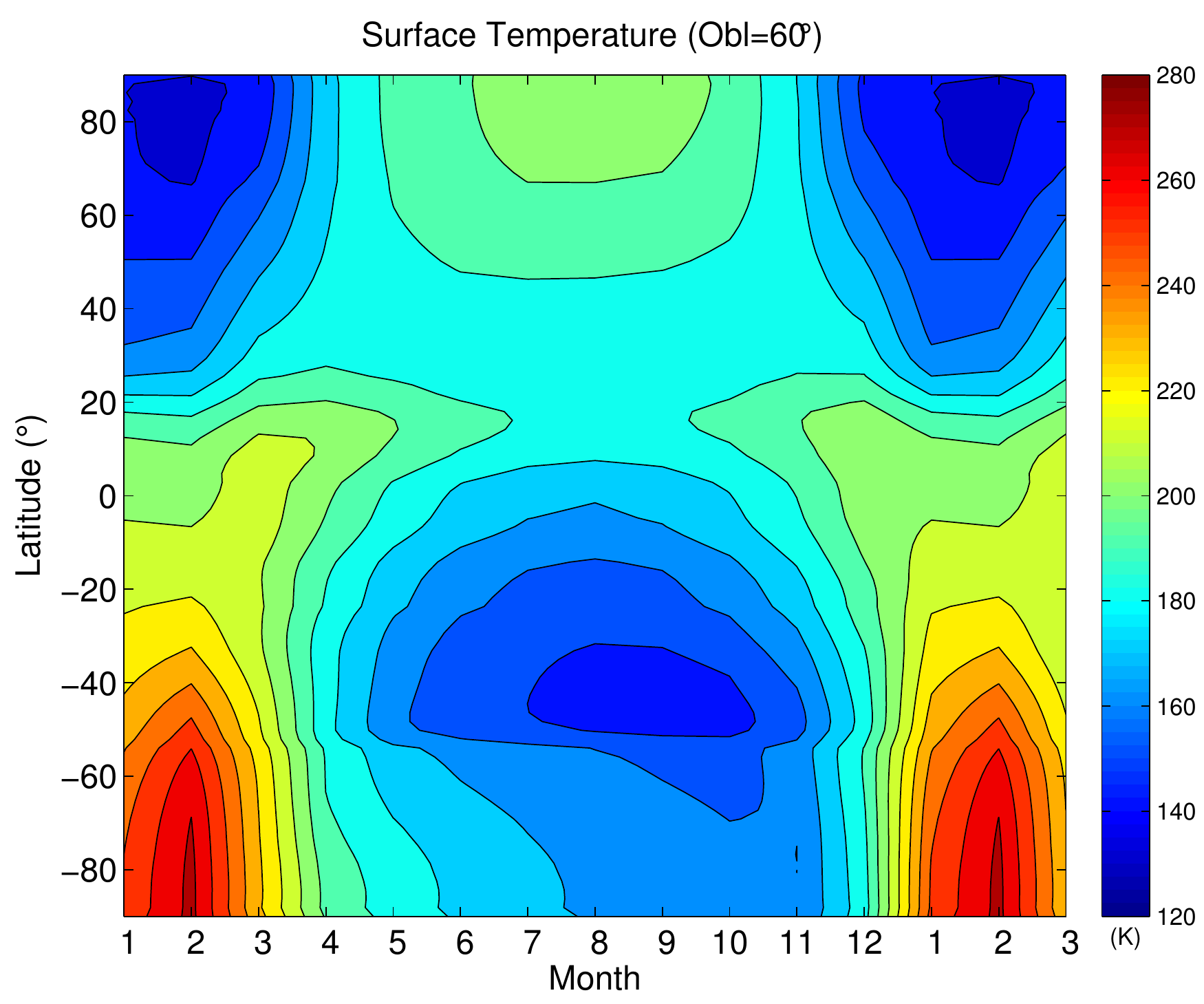}
\includegraphics [scale=0.4]{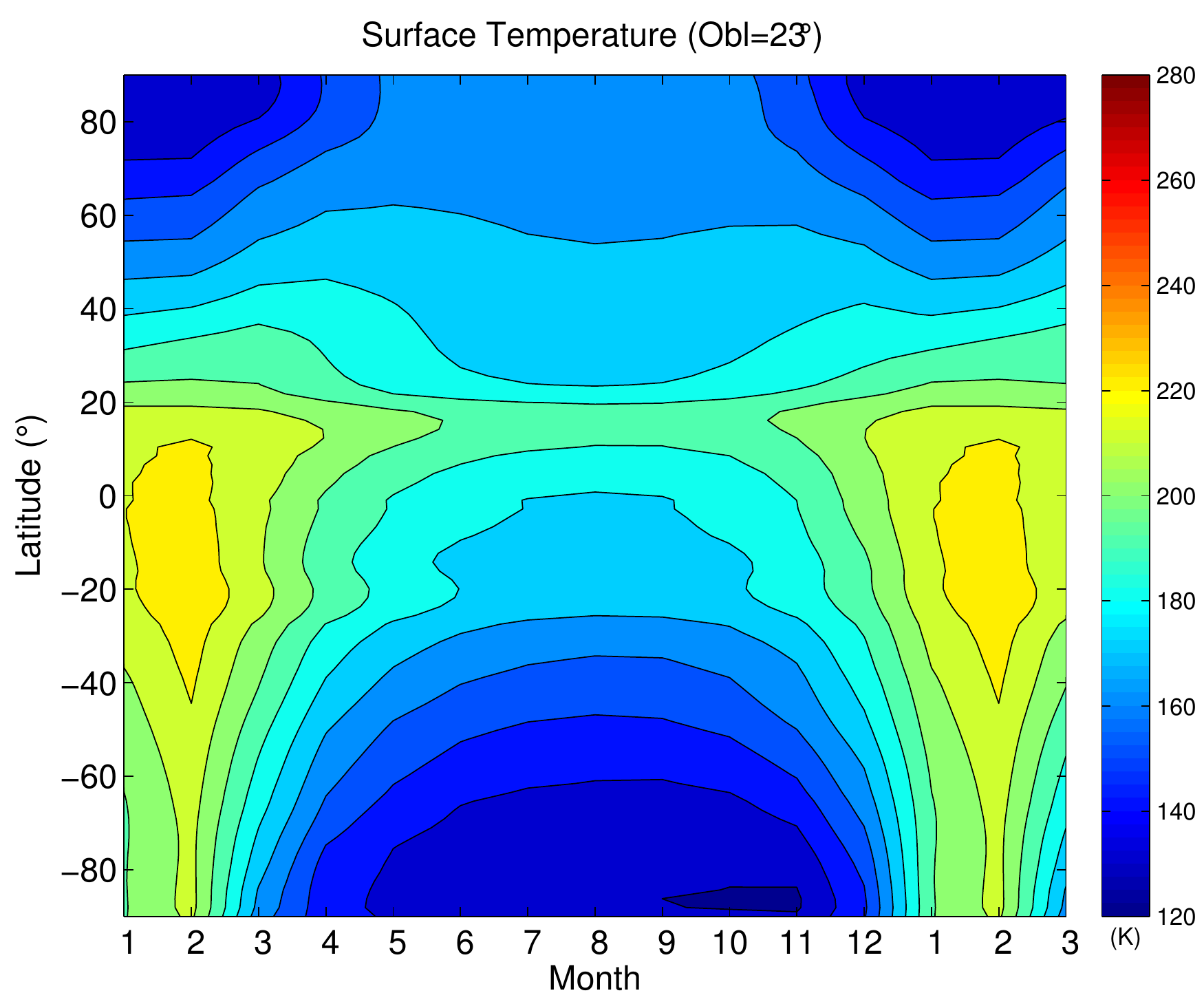}\\
\includegraphics [scale=0.4]{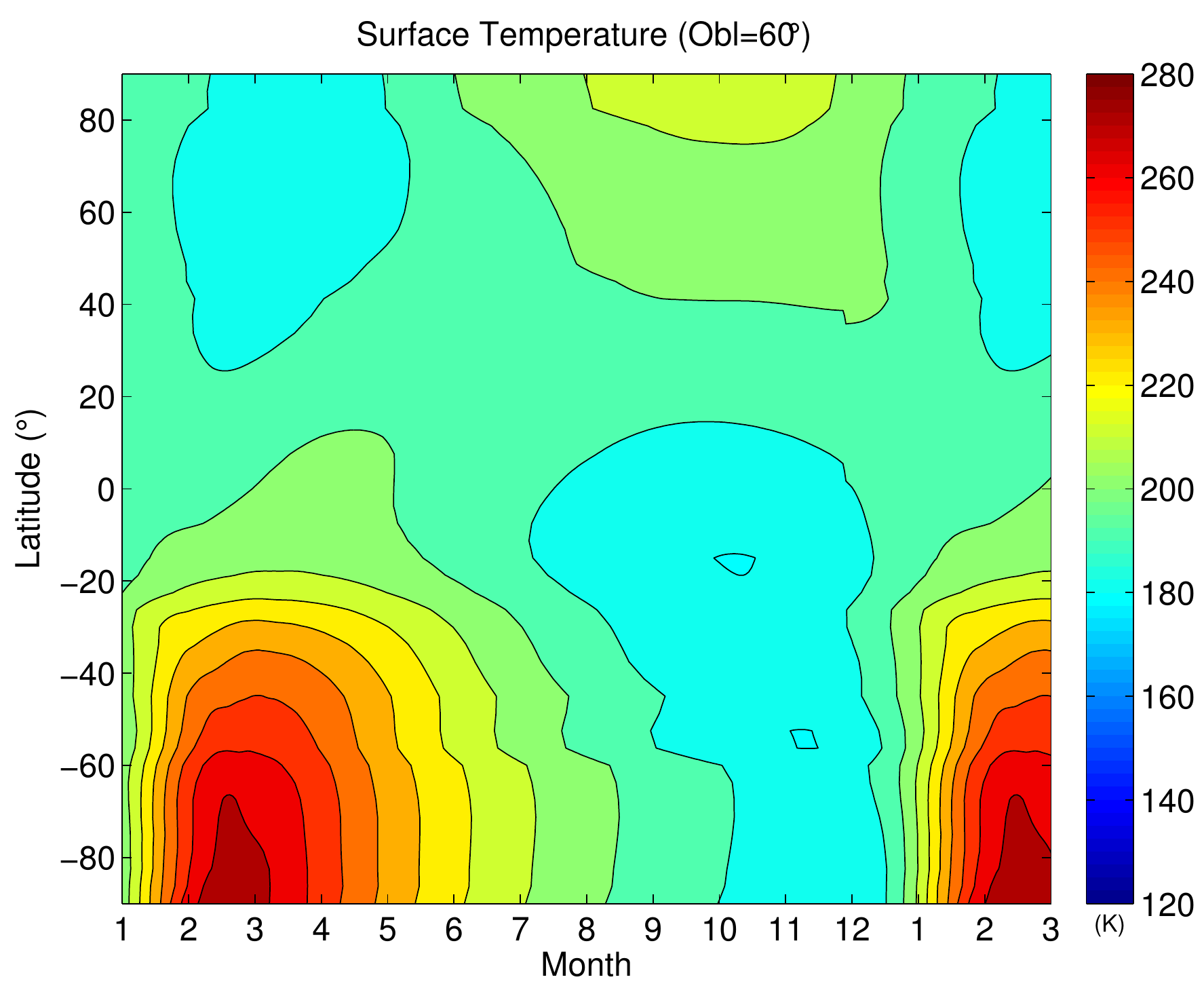}
\includegraphics [scale=0.4]{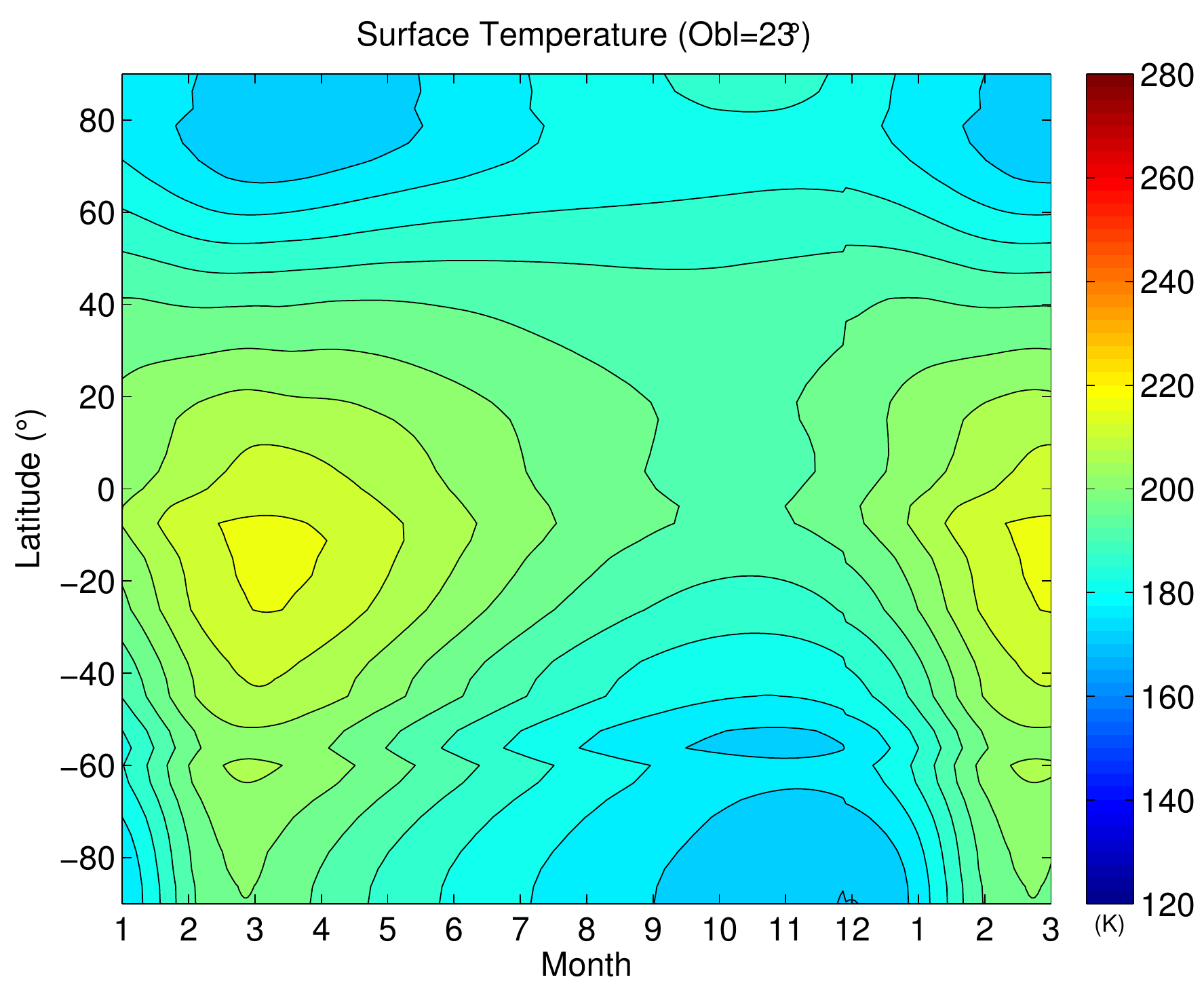}\\
\caption{Top: Annual mean instellation as a function of latitude for Kepler-62f after 40-yr CCSM4 simulations, assuming the present atmospheric level of CO$_2$ on Earth (400 ppmv), an obliquity of 23$^\circ$ (blue) and 60$^\circ$ (green). Middle: Surface temperature as a function of the month of the year, assuming a 12-month annual cycle, for an obliquity of 60$^\circ$ (left) and 23$^\circ$ (right), after 40-yr CCSM4 simulations. Bottom: Surface temperature as a function of month of year for an obliquity of 60$^\circ$ (left) and 23$^\circ$ (right), after 60-yr LMD Generic GCM simulations. In both models VEP was set to 90$^\circ$, similar to the Earth (102.7$^\circ$). The eccentricity was set to 0.32. The less extreme cold temperatures in the LMD Generic GCM simulations are due to a 10-m maximum thickness limit of sea ice in LMD Generic GCM, while the CCSM4 sea ice thickness near the poles is $\sim$30 m.}
\label{Figure 12.}
\end{center}
\end{figure}
\pagebreak

\begin{figure}[!htb]
\begin{center}
\includegraphics [scale=0.50]{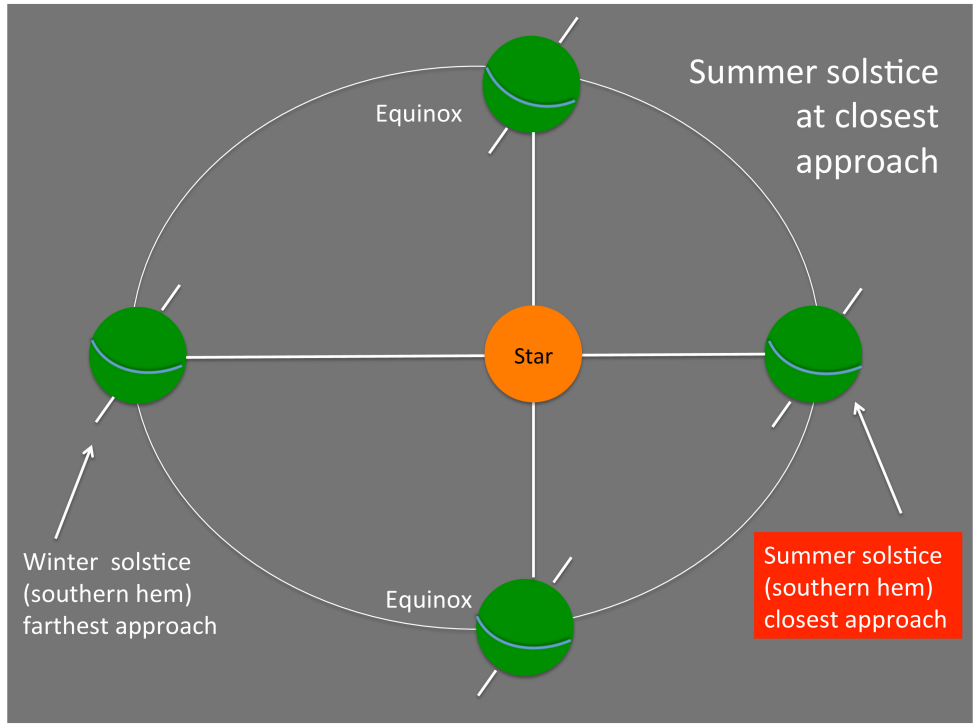}
\caption{Schematic diagram of assumed orbital configuration for CCSM4 (and select LMD Generic GCM) simulations of Kepler-62f. The angle of the vernal equinox with respect to pericenter was set to 90$^\circ$, similar to the Earth (102.7$^\circ$).}
\label{Figure 13.}
\end{center}
\end{figure}
\pagebreak

\begin{figure}[!htb]
\begin{center}
\includegraphics [scale=0.45]{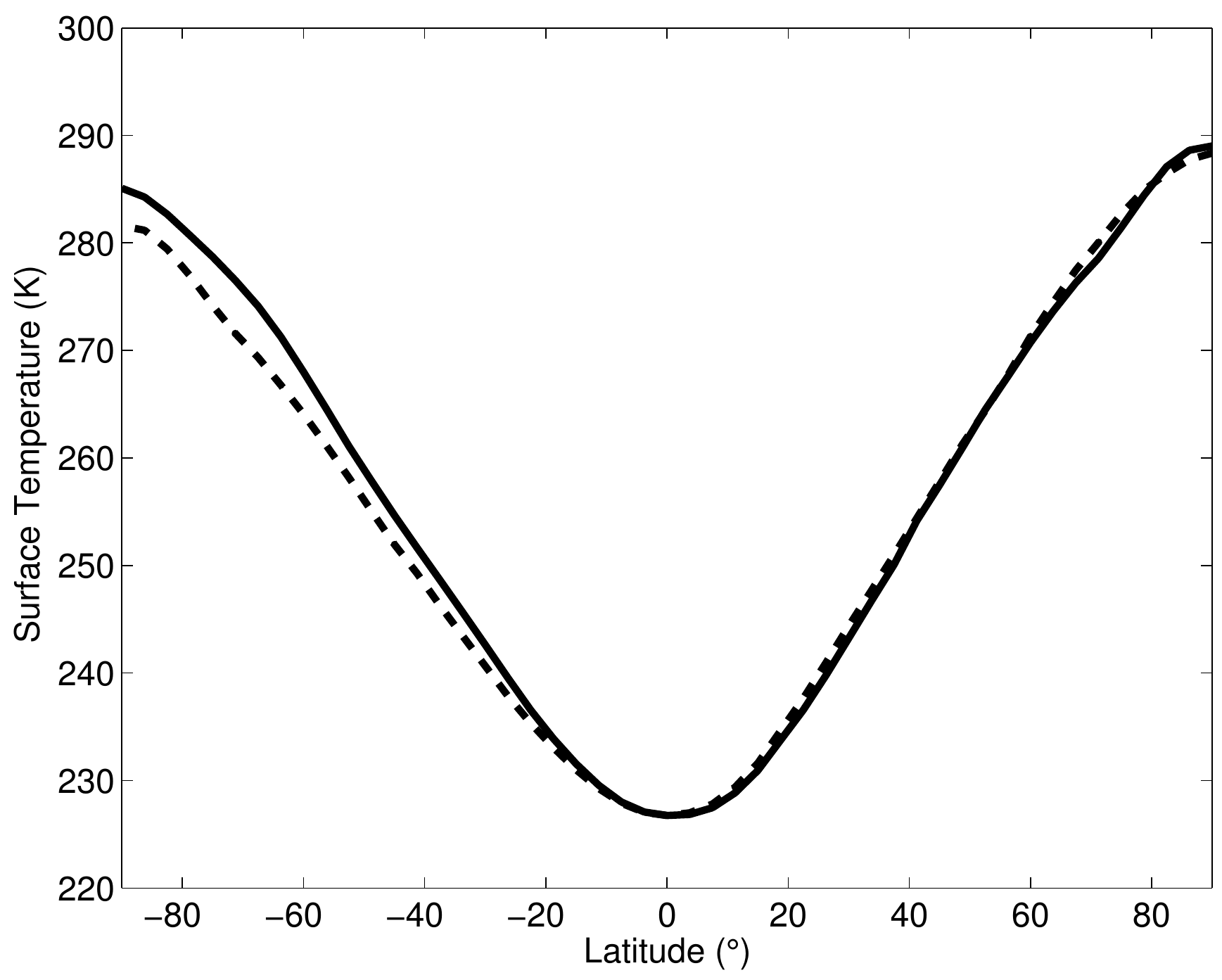}
\caption{Surface temperature as a function of latitude for Kepler-62f, after 160-year LMD Generic GCM simulations with 3 bars of atmospheric CO$_2$ and VEP is 0$^\circ$ (dashed line) and 90$^\circ$ (solid line). An obliquity of  90$^\circ$ and an eccentricity of 0.32 is assumed.}
\label{Figure 14.}
\end{center}
\end{figure}

\end{document}